\documentclass[useAMS,usenatbib]{mnras}
\usepackage{amsmath} 
\usepackage{graphicx}
\usepackage{float}

\title[Parameter Estimation for Stellar Populations]{Parameter Estimation for Scarce Stellar Populations}
          
\author[Ram\'irez-Siordia, V. H., Bruzual, G., Cervantes Sodi, B., and Bitsakis, T.]{V. H. Ram\'irez-Siordia$^{1}$\thanks{E-mail: manchasman@gmail.com},  G. Bruzual$^{1}$, B. Cervantes Sodi$^{1}$ and T. Bitsakis$^{1}$\thanks{CONACyT Research Fellow}\\$^{1}$Instituto de Radioastronom\'ia y Astrof\'isica, Universidad Nacional Aut\'onoma de M\'exico, Morelia, Michoac\'an, 58089 M\'exico}

\begin{document}
\date{Accepted . Received ; in original form }

\pagerange{\pageref{firstpage}--\pageref{lastpage}} \pubyear{2019}

\maketitle

\label{firstpage}

\begin{abstract}

We present a Bayesian method to determine {\it simultaneously} the age, metallicity, distance modulus, and interstellar reddening by dust of any resolved stellar population, by comparing the observed and synthetic color magnitude diagrams
 on a star by star basis, with no need to bin the data into a carefully selected magnitude grid.
We test the method with mock stellar populations, 
and show that it works correctly even for scarce stellar populations with only one or two hundred stars above the main sequence turn off. 
If the population is the result of two star formation bursts, we can infer the contribution of each event to the total stellar population. 
The code works automatically and has already been used to study massive amounts of Magellanic clouds  photometric data.
In this paper we analyze in detail three Large Magellanic Cloud star clusters and 6 Ultra Faint Dwarf Galaxies.
For these galaxies we recover physical parameters in agreement with those quoted in the literature, age 
$\sim13.7$ Gyr and a very low metallicity $log\,Z\sim-4$. Searching for multiple populations in these
galaxies, we find, at a very low significance level, signs of a double stellar population for Ursa Major I:
a dominant old population and a younger one which contributes $\sim25$\% of the stars, 
in agreement with independent results from other authors.

\,

\, 

\,

\,

\,

\,

\,

\,

\,

\end{abstract}


\section{Introduction}

The star formation history (SFH) of resolved stellar populations, like star clusters or dwarf galaxies, is usually derived from their colour-magnitude diagram (CMD).
The number of stars in characteristic phases of stellar evolution in the CMD carry information on the age, the metallicity, and the strength of the star formation burst. For example, the position of the Main Sequence Turn Off (MSTO) is sensitive to age, the presence of bright and blue mains sequence (MS) stars is indicative of a young star burst, Horizontal Branch (HB) and RR-Lyrae stars are characteristic of low-metallicity old stellar populations.
By fitting synthetic or mock CMDs computed for a wide range of physical parameters to an observed CMD, we can estimate the best age ($t$), metallicity ($Z$), distance modulus $(m-M)$, colour excess 
$E(\lambda_1-\lambda_2)$,
and stellar mass that describe a given stellar population, according to a particular set of stellar evolution models, within the observational errors.
Several methods have been developed to find the best match between observed CMDs and theoretical isochrones, ranging from the simple fit by eye, to more refined statistical techniques.
Fits by eye are subjective, do not provide confidence intervals for the estimated parameters, and become
extremely laborious when the task implies numerous stellar systems.\footnote{
For a review on the derivation of SFHs from CMDs see \cite{Gallart2005}.}

Fits can also be performed on the number of observed and expected stars inside conveniently chosen colour and magnitude bins in the CMD.
The observed and theoretical CMDs, which must include modeling of the photometric errors, are binned in identical fashion and the number of stars in corresponding bins are compared.
Minimizing $\chi^2$ is not the most convenient parameter estimator for poorly populated CMDs, since to obtain meaningful results, the number of stars in each bin must be statistically significant ($N \geq 10$), a task feasible in richly populated CMDs. Binning methods that overcome this limitation have been proposed, 
e.g., \cite{Mighell1999}, \cite{Dolphin2002}, and \cite{Aparicio2009}.

In this paper we develop a Bayesian inference code to estimate {\it simultaneously}, from {\it unbinned} CMDs,
the set of parameters $t, Z, (m-M)$, $E(\lambda_1-\lambda_2)$, and the SFH characterizing {\it single} or 
{\it double} stellar populations.
A similar technique was used by \cite{Tolstoy1996b} and \cite{Tolstoy1996} to determine the SFH of single
stellar populations, and by \cite{Hernandez2008} and \cite{Perren2015} to determine $t, Z, (m-M)$, and $E(\lambda_1-\lambda_2)$ for the same kind of populations.
Our approach differs from theirs in the way that we build the likelihood distribution function for all the parameters under consideration, and in the use of the marginalized probability distribution functions 
to assess the confidence intervals for the validity of our estimates.
More importantly, we show that our method can be used to estimate the parameters of double stellar
populations, including the stellar contribution of each burst, and examine the behavior of our results varying the sample size, its limiting magnitude, and the photometric errors in the CMD.
Our approach can be easily extended to study multiple ($> 2$) populations.

A first aim of this project is to estimate the parameters of a large number of young star clusters in
the Magellanic Clouds in an objective and automated manner. 
In this paper we study three LMC clusters in detail.
In \citet{Bitsakis2017,Bitsakis2018} we used our parameter inference tool to date systematically 4850 clusters in the Largel Magellanic Cloud (LMC) and 1319 clusters in the Small Magellanic Cloud (SMC), respectively.
These clusters range in age from 10 to a few hundred Myr and are sparsely populated, containing typically
$\sim 100$ stars. Many of them have not been catalogued previously. 
In this paper we study three LMC clusters in detail, two of them handpicked by inspecting their CMDs and
selecting those showing features that denote the presence of double stellar populations. Applying our method, we
are able to distinguish the presence of two stellar populations, establish their ages, and the stellar contribution of each burst.

A second goal of this paper is to revisit the derivation of the SFH and other physical parameters of a sample of six Ultra Faint Dwarf Galaxies (UFDGs) observed with the Hubble Space Telescope (HST) by 
\cite{Brown2012,Brown2014}. 
UFDGs are interesting because they are thought to be fossils from the first star bursts in the universe. They are indeed old, with ages $\sim13$ Gyr, show a very low metal content, $Z\sim0.0001$ \citep{Kirby2008,Brown2014},
and their SFHs seem to imply that a synchronized global event, such as reionization, deprived them of gas and quenched star formation \citep{Ricotti2005}. The number of resolved stars in a UFDG is typically  $\sim $1000, with only $\sim$ 100 of them above the MSTO. 
Different methods have been used to estimate their SFHs, from isochrone fitting by eye \citep{Sand2012} to adequate CMD binning methods \citep{Brown2014,Weisz2014}. \cite{Brown2014} adopt the maximum likelihoood binning method of \cite{Dolphin2002} (reviewed by \cite{Walmswell2013}), using a binning grid from the MSTO to the top of the Sub Giant Branch (SGB), excluding zones with few stars.
As far as we are aware, our analysis of the UFDGs is the first one to use most of the observed stars in
the CMD, on a star by star basis without binning the data, exploring the most sensitive areas of the CMD to reduce the confidence intervals of the estimated parameters.
 
In \S2 we describe in detail our implementation of the Bayesian inference approach.
In \S3 we apply our method to study the CMDs of single and double burst mock populations, recovering their SFHs and physical properties, and measure the capabilities and limitations of the method by performing a series of controlled tests. 
In \S4 we study a sample of three LMC star clusters and six UFDGs and compare our results with previous determinations. 
Our general conclusions are summarized in \S5.

\section{BAYESIAN INFERENCE}

\subsection{Isochrones as PDFs}\label{isochrones}

An isochrone is the theoretical locus in the CMD defined by stars of a given age and metallicity according
to a given set of stellar evolutionary tracks.
The shape of the isochrone in the CMD depends on the age ($t$) and metallicity ($Z$) of the stellar
population, and on the photometric bands in use. 
The isochrone must be shifted in the vertical (magnitude) axis by the distance modulus $(m-M)$ and the
extinction ($A_\lambda$) along the line of sight to the stellar system in study.
In the horizontal (colour) axis the isochrone is displaced according to the colour excess 
$E(\lambda_1-\lambda_2)$, which is related to $A_\lambda$ by the reddening law.
Without lack of generality, we write the $ith$-isochrone selected from a large set as the function
\begin{equation}\label{eq0}
ith-{\rm isochrone} = h_i[t, Z, m-M, A_\lambda, E(\lambda_1-\lambda_2)].
\end{equation} 
Stars formed in a single burst are expected to fall in the CMD along the locus corresponding to a specific isochrone.
Given our probabilistic approach, we treat $h_i$ as a probability distribution function (PDF): 
the probability of finding a star at a given position in the CMD is then proportional to the 
number of stars expected at this position according to the assumed stellar initial mass function (IMF).

\begin{figure}
\centering
\includegraphics[width=6.6cm]{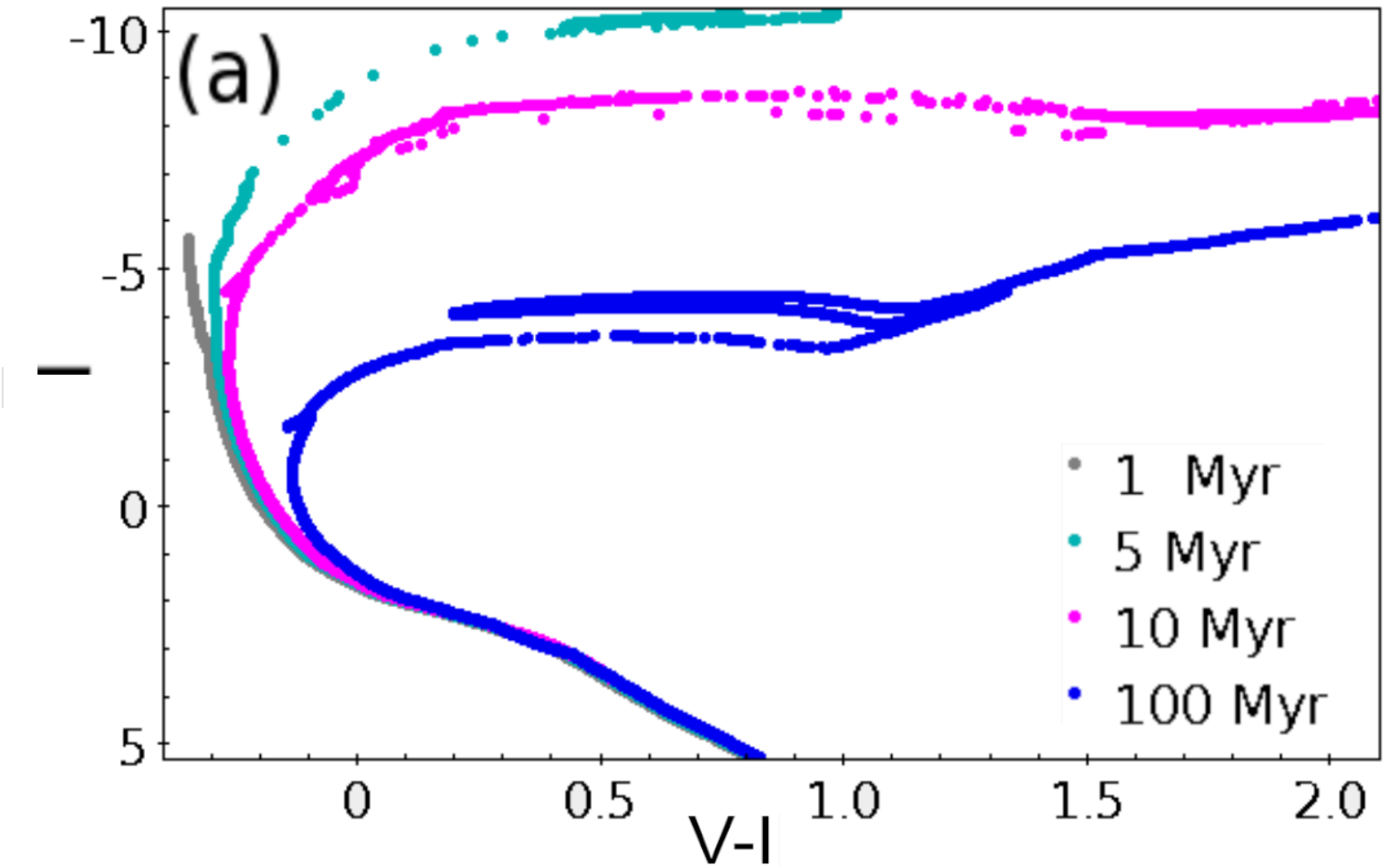}
\includegraphics[width=6.0cm]{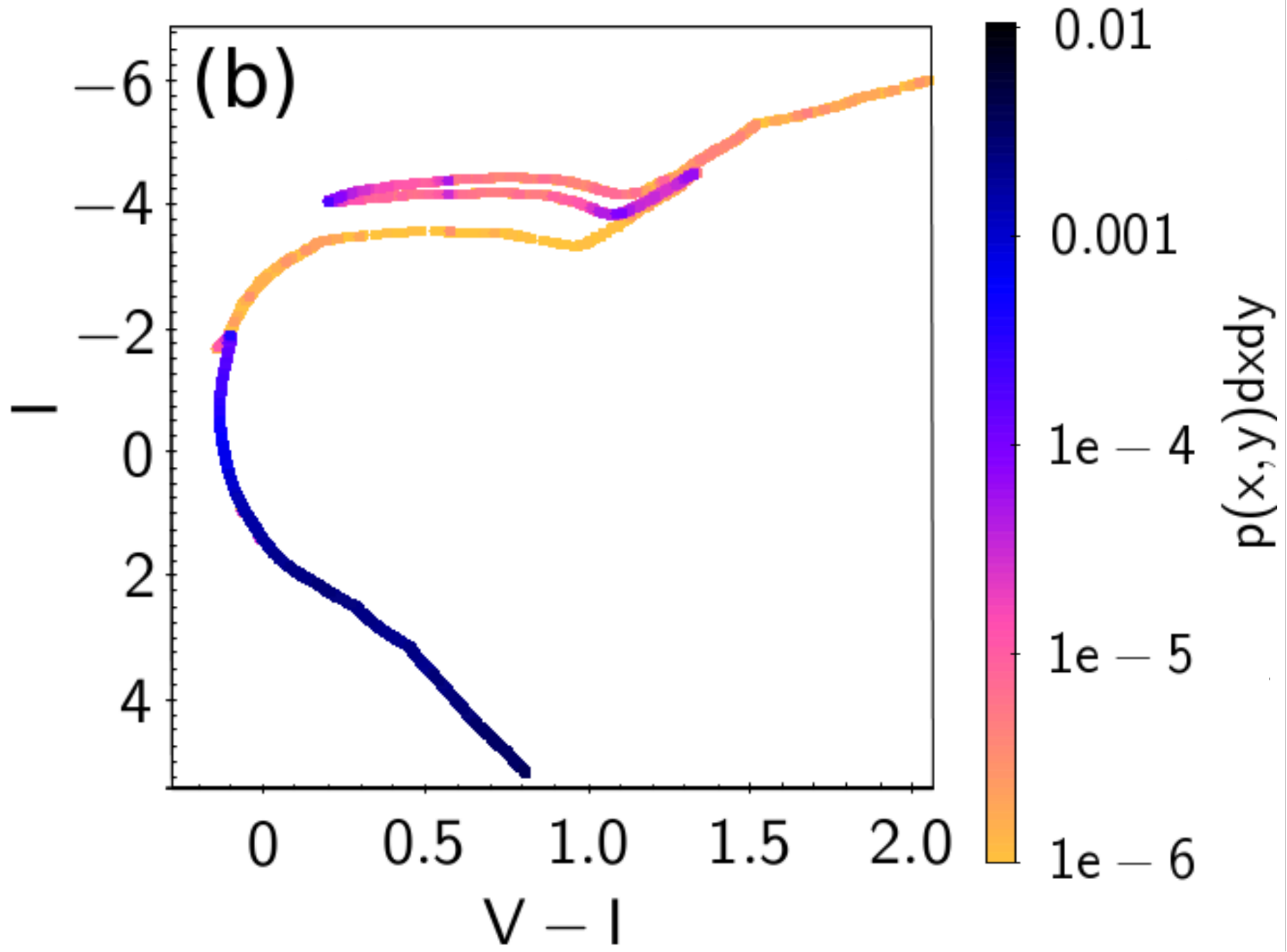}
\caption{(a) $Z = 0.008$ isochrones in the $I$ vs. $V-I$ CMD for age =  1, 5, 10, and 100 $Myr$.
As times goes by, the shape of the isochrone changes in the CMD.
(b) $Z = 0.008$, 100 $Myr$ isochrone as a PDF. The colour scale represents the relative probability of
finding a star at a given position, which depends on the time spent by the stars at the given evolutionary
phase. It is more likely to find a star on the MS than on the HB, the RGB, or the SGB.}
\label{fig1}
\end{figure}

In this work we use isochrones computed from the \cite{Chen2015} and \cite{Marigo2013} evolutionary tracks 
by Charlot \& Bruzual (in preparation) using the \cite{CharlotBruzual1991} isochrone synthesis algorithm.
Isochrones are available for 16 values of the metallicity\footnote{$Z=0.00005$, 0.0001, 0.0002, 0.0005, 0.001, 0.002, 0.004, 0.006, 0.008, 0.01, 0.014, 0.017, 0.02, 0.03, 0.04, and 0.06} at ages ranging from 
$10^4$ yr to $15$ Gyr in 220 time steps.
The BaSeL 3.1 atlas \citep{Westera2002} is used to derive the photometric properties of the stars.
The isochrones are populated stochastically as described by \citet{Bruzual2010} following the \citet{Kroupa2001} universal IMF assuming that stars form initially in the mass
range from $m_L = 0.10$ to $m_U = 100$ M$_\odot$. 
Isochrones must be densely populated to resemble as close as possible a continuous PDF in the CMD. 
By trial and error we found that this is achieved with a population of 1 to 2 million stars, warranting 
$\sim$ 100,000 stars above the MSTO, where the isochrones tend to be more sparsely populated due
to the rapid transit of the stars through the post MS evolutionary stages.
This is illustrated in Fig.~\ref{fig1}.

\subsection{Bayesian procedure}\label{algorithm}

To determine the age and physical properties of the stars in a stellar population, we use a Bayesian approach to infer the stellar population parameters that reproduce the observed CMD. A detailed analytic description on the association of a likelihood value to a given set of parameters is presented in \cite{Tolstoy1996}, and \cite{Hernandez1999}, and reviewed in \cite{Walmswell2013}.

Let $o$ represent an observed sample of stars in a CMD and $h$ a set of hypotheses (isochrones) which depend on the parameters in Eq.~(\ref{eq0}) 
The Bayes rule provides the posterior probability $p(h_{i}|o)$ for each hypothesis given the observation
\begin{equation}\label{eq1}
p(h_{i}|o)=\frac{p(o|h_{i})p(h_{i})}{p(o)},
\end{equation}
where $p(o|h_{i})$ is the probability of obtaining the observation given the $ith$-hypothesis, and $p(h_{i})$
and $p(o)$ are the probability of $h_i$ and the observation $o$, respectively.
The hypothesis $h_{max}$ that maximizes (\ref{eq1}) is known as the {\it estimate} and provides the most likely combination of parameters that reproduces the observations.
Finding the {\it estimate} is equivalent to estimating the free parameters, while the width of the likelihood distribution provides a measure of their confidence intervals. 
The quantity obtained by dividing Eq.~(\ref{eq1}) by $p(h_{max}|o)$ is referred to as the likelihood 
$\mathcal{L}(h_{i}|o)$,
\begin{equation}\label{eq2}
\mathcal{L}(h_{i}|o)=\frac{p(o|h_{i})p(h_{i})}{p(o|h_{max})p(h_{max})}.
\end{equation} 
If, as is our case, we lack prior information on $p(h_{i})$, all $p(h_{i})$ are considered equal, which 
simplifies Eq.~(\ref{eq2}) to
\begin{equation}\label{eq3}
\mathcal{L}(h_{i}|o)=\frac{p(o|h_{i})}{p(o|h_{max})}.
\end{equation} 
The resulting likelihood distribution, $\mathcal{L}(h_{i}|o)$, is not a PDF properly, but an indicator of the relative probability of occurrence of the hypothesis $h_i$.

To compute $p(o|h_{i})$ let us consider a single star $s_k$ observed at position $(x_{k},y_{k})$ in the CMD with observational errors $(\sigma_{x,k},\sigma_{y,k})$.
The probability that such star arises from a starburst characterized by hypothesis $h_{i}$ (defined by the $ith$ set of parameters), is a function of the position of the star weighted by the errors
\begin{equation}\label{eq4}
p(o|h_{i}) = p(s_{k}|h_{i}) = \int h_{i}(x,y)U_{k}(x-x_{k},y-y_{k})dxdy,
\end{equation}
where $U_{k}(x-x_{k},y-y_{k})$ is the error function for the $kth$ star, which we take to be the bivariate
Gaussian
\begin{equation}\label{eq5}
U_{k}(x-x_{k},y-y_{k}) = C_{k,x,y}\,\, e^{-\left(\frac{(x-x_{k})^{2}}{2\sigma_{x,k}^{2}}+\frac{(y-y_{k})^{2}}{2\sigma_{y,k}^{2}}\right)},
\end{equation}
where
\begin{equation}\label{eq6}
C_{k,x,y} = \frac{1}{2\pi\sigma_{x,k}\sigma_{y,k}}.
\end{equation}
Our isochrones are discrete sets of $n$ stars. If $(x_j,y_j)$ is the position in the CMD of the $jth$ star in the isochrone, then Eq.~(\ref{eq4}) can be expressed as the sum
\begin{equation}\label{eq7}
p(s_k|h_{i})=\frac{C_{k,x,y}}{n}\,\,\sum_{j=1}^{n}\exp{\left(-\frac{(x_j-x_{k})^{2}}{2\sigma_{x,k}^{2}}-\frac{(y_j-y_{k})^{2}}{2\sigma_{y,k}^{2}}\right)},
\end{equation}
where $n$ is the number of stars in the $ith$ isochrone, and $1/n$ corresponds to the normalization constant for this isochrone. From Eq.~(\ref{eq7}) we see that $p(s_k|h_{i})$ is defined if, and only if,
$\sigma_{x,k} > 0$ and $\sigma_{y,k} > 0$, i.e., non zero observational errors must be associated with each star $s_k$.

In most cases, observed CMDs contain a large number $N$ of stars.
Following \cite{Tolstoy1996}, we write the probability that the $N$ stars in the CMD come from a burst of star formation characterized by the hypothesis $h_i$ as
\begin{equation}\label{eq8}
\mathcal{L}_N(h_{i}|o) = \prod_{k=1}^{N}p(s_k|h_{i})^{1/N},
\end{equation}
and in logarithmic form as
\begin{equation}\label{eq9}
\ln \mathcal{L}_N(h_{i}|o) = \frac{1}{N}\ \sum_{k=1}^{N}\ln p(s_{k}|h_{i}).
\end{equation}
Evaluating the sum in Eq.~(\ref{eq9}) for the entire set of hypothesis $h_i$ we obtain the likelihood distribution function $\mathcal{L}_N$.\footnote{In practice, we compute $\mathcal{L}_N$ from Eq.~(\ref{eq9}) and divide the resulting distribution by its maximum, resulting in $\mathcal{L}_N^{max}=1$.}
From $\mathcal{L}_N$ we obtain its statistical properties, such as its mode, median, maximum, standard deviation, FWHM, skewness, etc. For a single parameter problem, the mode or median and the FWHM of $\mathcal{L}_N(\phi)$ provide reasonable estimates of the value of the parameter $\phi$ and its confidence interval $\sigma_\phi$. In the case of multi-parameter problems, estimating the value of each parameter and its confidence interval is more cumbersome, and using different schemes may lead to different answers \citep{Basu1977,Pawitan2013}. In this paper we use the marginalized likelihood distribution, $\mathcal{L}_N^{marg}(\phi_{q})$, to estimate the value of a parameter $\phi_q$.
$\mathcal{L}_N^{marg}(\phi_{q})$ is commonly used in Bayesian inference studies when there is no information about the nuisance parameters, which are assumed to follow flat probability density functions \citep{Verde2010,Pawitan2013}.
$\mathcal{L}_N^{marg}$ for $\phi_q$ is defined as:
\begin{equation}\label{eqMarginal}
\mathcal{L}^{marg}_N(\phi_{q})=\int\mathcal{L}(\phi_{q},\nu)d\nu,
\end{equation}
where $\nu$ refers to the set of nuisance parameters, i.e., all parameters except $\phi_q$ \citep{Verde2010}\footnote{We divide the resulting distribution by its maximum, resulting in $\mathcal{L}^{marg}_N(\phi_{q})^{max}=1$.}. The mode of the $\mathcal{L}_N^{marg}$ distribution leads to the estimate $\phi_{q}^{marg}$, the most likely value of the parameter that integrates all the possibilities of the nuisance parameters. The confidence interval $\sigma^{marg}_q$ is obtained from the half maximum of $\mathcal{L}_N^{marg}$.

If a problem stellar population arises from a double burst of star formation, we evaluate $\mathcal{L}_N$ for all possible linear combinations of pairs $(i,j)$ of isochrones in our hypothesis set, namely
\begin{equation}\label{eq10}
h_{i,j,w}=w\,h_{i}(x,y) + (1-w)\,h_{j}(x,y),
\end{equation}
where the weights $w$, and $1-w$, account for the fraction of stars belonging to the $ith$ and the $jth$ isochrones, respectively. The formalism in Eq.~(\ref{eq10}) can be extended to multiple stellar populations, i.e, more than two stellar bursts \citep{Dolphin1997}. Each isochrone in (\ref{eq10}) corresponds to a different star formation episode.

\subsection{Parameters and priors}\label{ParametrosYPriors}

From the procedure described above, we estimate four basic parameters
characterizing a stellar population $(t, Z, m-M, E(\lambda_1-\lambda_2))$. 
For a complete sample of observed stars, we can estimate the total mass of the stellar population.
In the case of a double burst population described by Eq.~(\ref{eq10}), we can recover, in principle,
seven parameters ($t_1, t_2, Z_1, Z_2, m-M, E(\lambda_1-\lambda_2),  w_1/w_2$).

The number of free parameters to estimate depends on our prior knowledge on them. Using prior information reduces computational time and resources, and may help to break degeneracies in the solution.
For instance, prior information about the metallicity of the stellar population, e.g., $Z\pm \sigma_Z$, can be adopted to break the age-metallicity degeneracy. The uncertainty $\sigma_Z$ is introduced in the analysis by marginalizing over the isochrone sets that cover the range of possible values of $Z$. As a rule, in this work we marginalize over isochrone sets expanding a range of $\pm 5 \sigma_Z$ (see \S\ref{LMC}).

Prior knowledge on the values of $(m-M)$ and/or $E(\lambda_1-\lambda_2)$, can be adopted to solve the problem only for $t$ and $Z$. In this case, the associated uncertainties, $\sigma_{(m-M)}$ and $\sigma_{E(\lambda_1-\lambda_2)}$, should be added in quadrature to the photometric uncertainties of the stars in the CMD, i.e.,

\begin{subequations}\label{eq12}
\begin{align}
\sigma_{x} &= \sqrt{\sigma_{\lambda_1-\lambda_2}^2  + \sigma_{E(\lambda_1-\lambda_2)}^2 + \sigma_{S_x}^2},\\
\sigma_{y} &= \sqrt{\sigma_{\lambda_2}^2  + \sigma_{(m-M)}^2 + \sigma_{S_y}^2}, 
\end{align}
\end{subequations}
where $\sigma_{\lambda_1-\lambda_2}$ and $\sigma_{\lambda_2}$, are the photometric uncertainties on the x and y-axis, respectively (see \S\ref{LMC}). $\sigma_{S_x}$ and $\sigma_{S_y}$ correspond to the systematic uncertainties.

In absence of previous (or not adequate) measurements of a nuisance parameter, we can estimate its value using our procedure by setting it as a free parameter. Often we have information that can be used 
to delimit the range of possible values of this parameter, e.g., when analyzing stellar populations in the Milky Way, we constrain the domain of $(m-M)$ to Galactic scales. 
Delimiting properly the domain of a free parameter is important since the marginalized distributions may
depend on the chosen domain, as happens with any prior \citep{Verde2010}.
In our case, when a previous measurement of a nuisance parameter is available
(let's say $\phi=\phi_0\pm\sigma_0$), 
and we decide to estimate its value, we choose its domain as: $[\phi_0-5\sigma_0,\,\phi_0+5\sigma_0]$.

\subsection{Advantages of using unbinned CMDs}\label{unbinning}

Our method compares model and observed CMDs on a star by star basis, and does not require a grid of binned data, in contrast to $\chi^2$ or Poisson based statistics \citep{Harris2001,Dolphin2002}, avoiding the dilemma of arbitrarily defining the size, form, and position of the bins in the CMD.
Bins should be small enough to resolve sensitive stellar phases on each CMD, but large enough to contain a significant number of stars.
Data binning degrades the CMD resolution, results in a lost of valuable information, may introduce unnecessary noise and statistical errors, and compromises the effectiveness of these methods for the study of scarce stellar populations, as pointed out by \cite{Walmswell2013}. 
Badly chosen bins may affect the results of $\chi^2$ based fits, and may introduce undesired biases and uncertainties.
We avoid rejecting valuable information from stars in poorly filled bins, which allow us to estimate the physical parameters of scarce stellar populations.  
Using unbinned data is also advantageous in the context of processing automatically and blindly catalogues 
containing large numbers of CMDs,\footnote{
In general, these catalogues are inhomogeneous, and several properties of the CMDs
which are critical in designing an appropriate grid of bins,
such as the number of stars, their dispersion in position due to the photometric uncertainties, the overall morphology due to different age or reddening, or the possibility that the catalogue mixes data from different observational programs, vary from cluster to cluster.}
as has been shown by \cite{Bitsakis2017,Bitsakis2018} using our code on LMC and SMC star cluster catalogues.
A brief review on binned and unbinned methods can be found in \cite{Walmswell2013}.

\subsection{Trimming the sample}\label{trimming}

Some regions in the CMD are more sensitive to differences in age than others.
In general, the brighter the star, the more information it will contribute to unveiling the 
population parameters under analysis.
The differences in the isochrones are more pronounced around the MSTO and the SGB. 
In contrast, the isochrones overlap towards fainter magnitudes reaching a point where they become
indistinguishable (cf. Fig.~\ref{fig1}). 
Therefore, observed stars in the lower part of the MS do not contribute much to our analysis, and since 
the observational uncertainties increase towards the bottom of the CMD, including lower MS stars only 
adds noise to our results. In the next section we show, by applying different magnitude cuts to the sample, that including fainter stars does not
change the values estimated for the different parameters, only broadens the confidence intervals.
Thus, to lower the uncertainties in the parameter estimates, lower MS stars are usually removed from the sample.

\section{Mock Galaxies}

To test the capabilities and limitations of the parameter estimator algorithm described in \S\ref{algorithm}, we generate a series of mock stellar populations and then use the estimator to recover the input values 
of ($t, Z, m-M, E(\lambda_1-\lambda_2)$) for these synthetic observations.
Each mock galaxy is generated as described in \S\ref{isochrones} and its population is left to evolve up to
the desired age, that we choose arbitrarily for each experiment.
To produce a realistic ($V-I,I$) CMD we add random photometric Gaussian uncertainties to the $V$ and $I$
magnitude of each star on the isochrone.
For the mock galaxies discussed in this section we model the photometric errors as a function of apparent magnitude using the relations
\begin{subequations}\label{eq13}
\begin{align}
\sigma_{V} & = 1.9\times10^{-8}\exp(0.5V),\\
\sigma_{I}  & = 4.7\times10^{-8}\exp(0.49I),\\
\sigma_{V-I} & = \sqrt{(\sigma_V)^{2}+(\sigma_{I})^{2}}.
\end{align}
\end{subequations}
derived from fits to the photometric uncertainties of the UFDGs observed with the HST by \cite{Brown2014}.\footnote{
In reality, the HST observations were performed in the $F606w$ and $F814w$ bands. For the purpose of the tests discussed in this section, we adopt for the $V$ and $I$ bands the photometric uncertainties characteristic of the $F606w$ and $F814w$ bands, respectively.} 
We allowed for a systematic uncertainty of $\sigma_S=0.02$ by adding it in quadrature to the photometric uncertainties before performing the analysis (Eq.~\ref{eq12}).

\subsection{Mock galaxy with a single stellar population}\label{mock1}

\begin{figure}

\includegraphics[width=8.45cm,height=12.1cm]{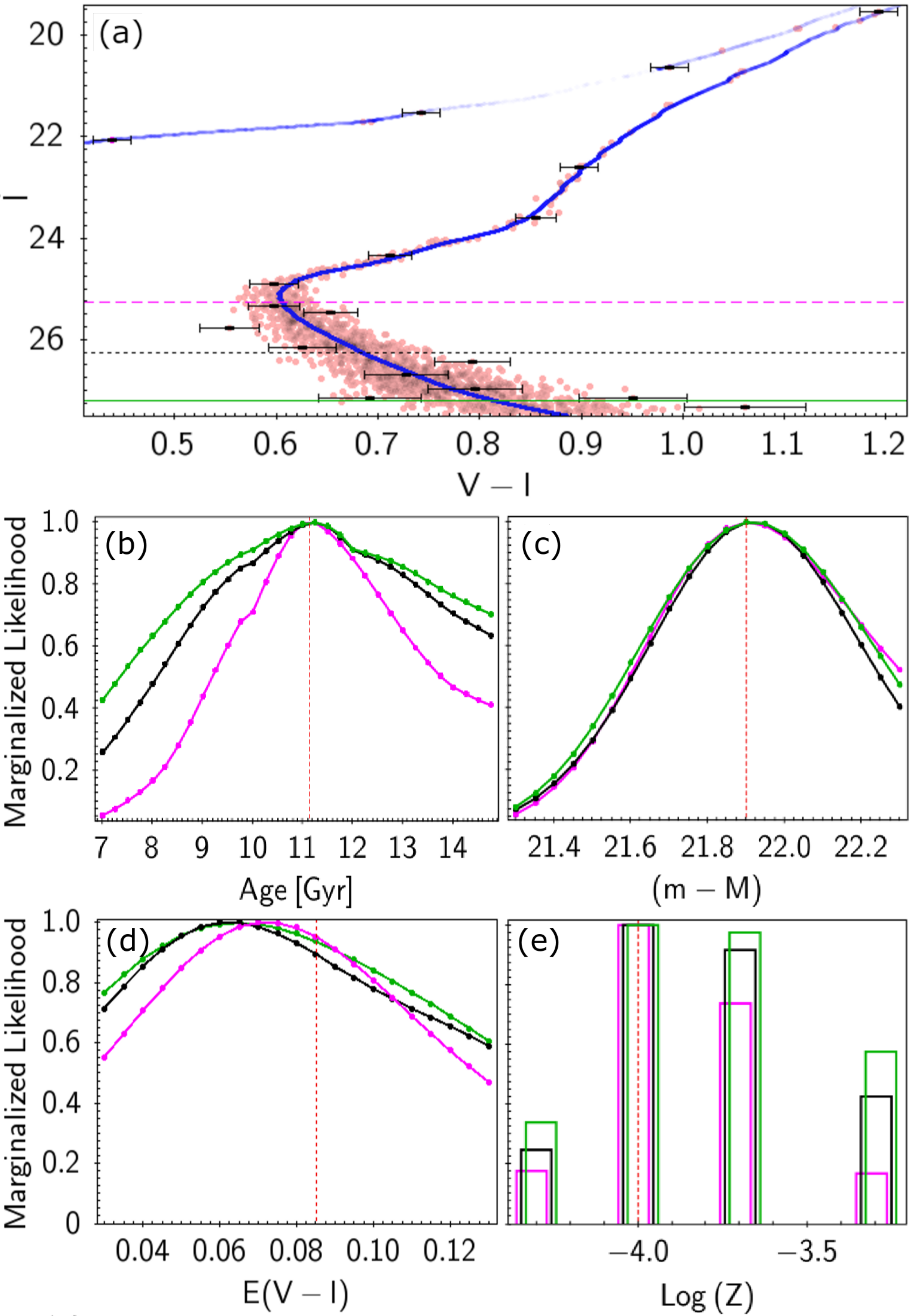}

\caption{
(a) CMD of a 11.125 Gyr old, $Log\, Z = -4$, mock simple stellar population containing 1608 stars, 160 above the MSTO. An input isochrone with its generating parameters (11.25 Gyr, $(m-M)=21.9$, $E(V-I)=0.085$, $Log\, Z=-4$) is shown in {\it blue}. Photometric uncertainties are shown for some stars. Horizontal lines mark three cuts on $I$ applied to the sample to test our algorithm. (b, c, d, e) $\mathcal{L}_N^{marg}$ distributions for $age$, $(m-M)$, $E(V-I)$, and $Z$. The 
{\it magenta, black}, and {\it green} lines correspond to the three different samples defined using the threes different cuts at mag 25.25, 26.25, and 27.25, respectively. 
The dotted vertical lines indicate the input parameters values.
}

\label{fig2}
\end{figure}

\begin{figure*}
\includegraphics[width=17.7cm,height=11.5cm]{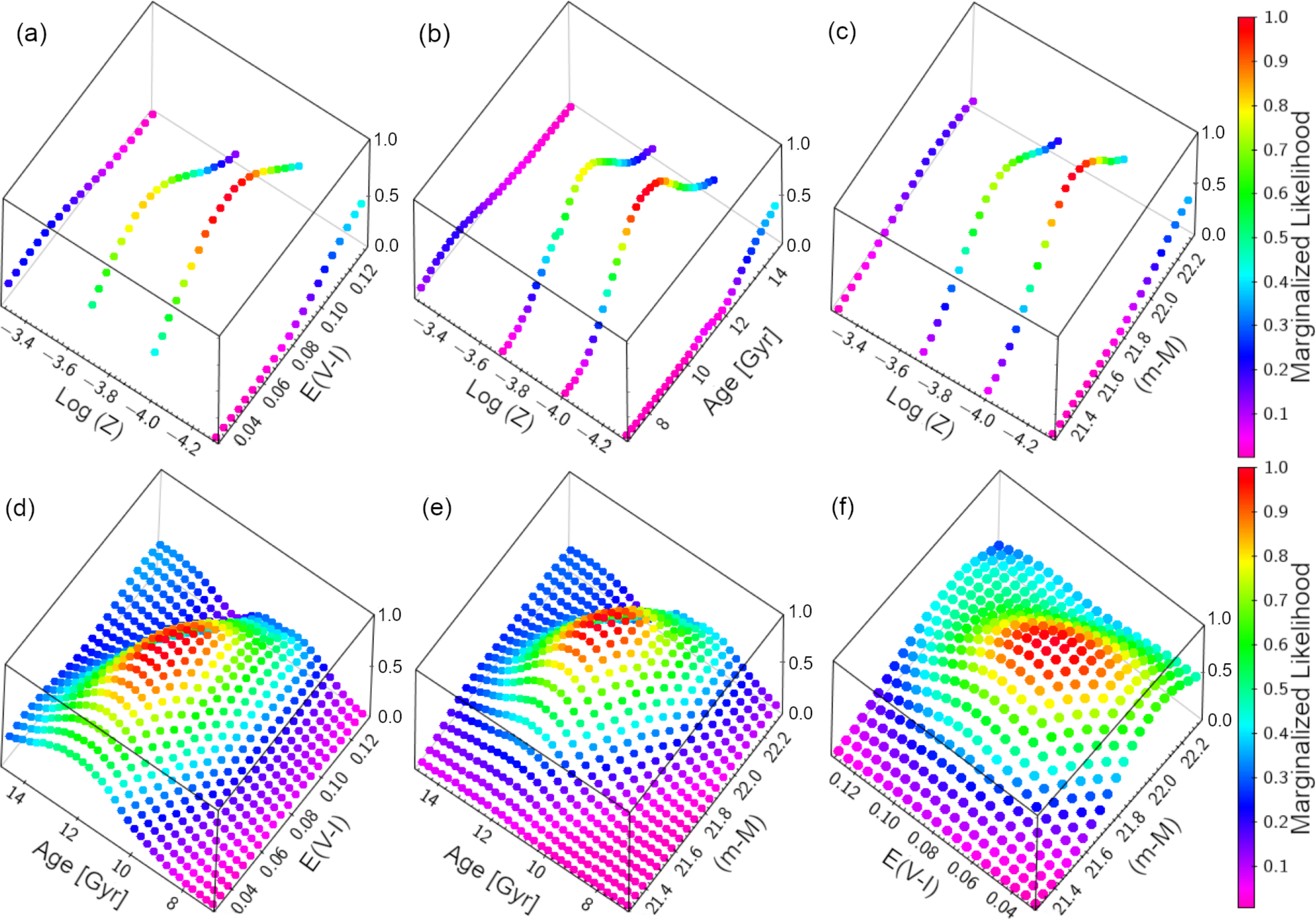}

\caption{2D $\mathcal{L}_N^{marg}$ distributions for all parameter pairs for the mock old simple stellar population of Fig~\ref{fig2}a.
}
\label{figA1}
\end{figure*}

Fig.~\ref{fig2}a shows the ($V-I,I$) CMD of a 11.125 Gyr old, $Log\,Z = -4$ mock simple stellar population at $(m-M) = 21.9$ with $E(V-I) = 0.085$. To make the test more meaningful, we choose an age that is not in our isochrone grid, adding in equal numbers of stars belonging to the 11 and 11.25 Gyr isochrones\footnote{
We use the same combination of ages in all the mock bursts considered in this paper.}
The resulting mock galaxy contains 1608 stars, only 160 of them above the MSTO, resembling an old dwarf 
satellite galaxy of the Milky Way. We derive $\mathcal{L}$ using stars brighter than three different magnitude cuts in the CMD, namely $I = 25.25$, 26.25, and 27.25.
Figs.~\ref{fig2}b-e show the resulting marginalized likelihood distributions $\mathcal{L}_N^{marg}$. We recover closely the input value of three of the free parameters ($age$, $(m-M)$, and $Z$), while $E(V-I)$ is slightly shifted to lower values. Estimates on $age$, $(m-M)$, and $Z$ do not change by including fainter stars in the analysis, but $\mathcal{L}_N^{marg}$ for $E(V-I)$ does change, shifting slightly towards lower reddening values. This effect is mainly due to the reddening-metallicity degeneracy. Fig.~\ref{figA1} shows the 2D $\mathcal{L}_N^{marg}$ distributions for all pair of fitted parameters. 
The partial relative contribution to the $E(V-I)$ $\mathcal{L}_N^{marg}$ by stellar populations of different
$Z$ can be seen in Fig.~\ref{figA1}a. Higher $Z$ values lead to lower reddening values. These figures  clearly show that degeneracies may dominate the inference results, producing non-unique solutions.\footnote{
A detailed study of the degeneracies in isochrone fitting in the context of a maximum likelihood approach based on Monte Carlo methods can be found in \cite{Hernandez2008}.}

Fig.~\ref{fig2} shows that as we include fainter stars, all the $\mathcal{L}_N^{marg}$ distributions broaden, indicating that
no meaningful information is added by including stars far below the MSTO, as discussed in \S\ref{trimming}. At first sight, this effect may seem puzzling, since one would expect that increasing the sample should 
result in reduced uncertainties. Here we note that as we include fainter stars, the proportion of stars with large photometric uncertainties increases dramatically. The fainter stars are located in zones of the CMD where the isochrones tend to be identical (cf. Fig.~\ref{fig1}); thus, all isochrones are equally able to recreate these faint stars, adding to the same extent to each isochrone's likelihood while reducing the relative weight from the few bright stars in regions of the CMD where the isochrones clearly
separate (around and above the MSTO), leading to wider likelihood distributions.
We adopt as our best solution the $\mathcal{L}_N^{marg}$ distributions for stars brighter than $I = 25.25$, resulting in the parameters listed in Table~\ref{tab1}.

\begin{table}
\centering
\caption{Mock galaxy with a single stellar population}
\label{tab1}
\begin{tabular}{|c|c|c|}
\hline 
 Parameter & Input value & Marg Estimate  \\
\hline 
 $t$ [Gyr] & $11.125$  &$11.25_{-2.0}^{+2.75}$         \\
 $Log (Z)$ &  -4       &$-4_{-0.3}^{+0.4}    $        \\
 $(m-M)$   & 21.9      &$21.9_{-0.3}^{+0.4}  $        \\
 $E(V-I)$  & 0.085     &    $0.075\pm0.05    $         \\
\hline 
\end{tabular}
\end{table}

\subsection{Mock galaxy with two stellar populations}\label{mock2}

\subsubsection{Two old stellar populations}\label{mock2b}
\begin{table}
\centering
\caption{Mock galaxy with 2 old stellar populations}
\label{tab2}
\begin{tabular}{|c|c|c|}
\hline 
Parameter & Input value & Marg Estimate \\
\hline 
$t_{y}$ [Gyr] & $8.125$   & $8.0_{-3.2}^{+2}   $  \\
 $t_{o}$ [Gyr] & $12.125$  & $13.25_{-2.5}^{+3.8}$  \\
 $Log\,Z$      & -4.0      &  $-4 \pm0.5$ \\ 
 $w_{y}$       & 0.5       & $0.5\pm0.35$  \\
 $(m-M)$       & 21.9      & $21.9\pm0.3$  \\ 
 $E(V-I)$      & 0.085     & $0.07\pm0.05$  \\

\hline 
\end{tabular}
\end{table}

\begin{table*}
\caption{Estimates from $\mathcal{L}_N^{marg}$ distributions for a mock galaxy with 2 old stellar populations of ages $t_{y}=8.125$ Gyr, $t_{o}=12.125$ Gyr, $Log\,Z=-4$, $(m-M)=21.9$, $E(V-I)=0.085$, and   varying $w_y$.
}
\label{tab3}
\begin{tabular}{|c|c|c|c|c|c|c|}
\hline
$w_{y,input}$    & $w_{y,est}$          & $t_{y,est}$ [Gyr]     & $t_{o,est}$ [Gyr]    &  $Log\,Z_{est}$  & $(m-M)_{est}$   & $E(V-I)_{est}$     \\ \hline
0.1             & $0.25_{-0.25}^{+0.4}$ &  $9.25_{-4.3}^{+3.7}$  & $13.75_{-2.4}^{+4.0}$ &  $-4 _{-0.3}^{+0.5}$    & 21.9$\pm$0.3    & $0.07_{-0.5}^{+0.7}$      \\
0.3             & $0.4_{-0.3}^{+0.35}$   &  $8.25_{-2.5}^{+2.1}$    & $13.5_{-2.0}^{+4.3}$ &$-4 _{-0.3}^{+0.5}$  & 21.9$\pm$0.3     & $0.07\pm0.05$     \\
0.5             & $0.5\pm0.35$       &  $8.0_{-3.2}^{+2}$          & $13.25_{-2.8}^{+4.5}$ & $-4 \pm0.5$         & $21.9\pm0.3$    & $0.07\pm0.05$   \\
0.7             & $0.65_{-0.35}^{+0.3}$      & $8.0_{-2.0}^{+1.5}$      &  $13.25_{-2.8}^{+4.5}$ & $-4 _{-0.3}^{+0.5}$    & 21.9$\pm$0.25    & $0.07_{-0.6}^{+0.7}$      \\
0.9             & $0.75_{-0.45}^{+0.25}$       &   $7.75_{-2.0}^{+1.5}$   &  $12.0_{-4.0}^{+5.5}$      &  $-4 \pm0.5$  & 21.9$\pm$0.25    & 0.07$\pm0.07$      \\ \hline

\end{tabular}
\end{table*}

\begin{table*}
\caption{Estimates from $\mathcal{L}_N^{marg}$ distributions for a mock galaxy with 2 young stellar populations of $Log\,Z=-2.1$, with $t_{y}=4.05$ Myr, $t_{o}=6.12$ Myr,  $(m-M)=29.0$, $E(V-I)=0.15$, and   varying $w_y$.}
\label{tab4}
\begin{tabular}{|l|l|l|l|l|l|}
\hline
$w_{y,input}$    & $w_{y,est}$  & $t_{y,est}$ [Myr]  & $t_{o,est}$ [Myr]  & $(m-M)_{est}$     & $E(V-I)_{est}$ \\ \hline
0.1             &       -       &           -         & $6.0_{-0.3}^{+1.8}$ & 29.0$\pm$0.3     & 0.15$\pm$0.025     \\
0.3             & $0.4_{-0.3}^{+0.4}$   & $4.2_{-0.4}^{+0.9 }$ & $6.8_{-1.1}^{+1.0}$        & 28.9$\pm$0.3    & 0.145$\pm$0.025      \\
0.5             & $0.5_{-0.35}^{+0.35}$   & $4.0_{-0.2}^{+1.0}$& 6.8$\pm$1.1        & $29.0_{-0.4}^{+0.2}$     & 0.145$\pm$0.025       \\
0.7             & $0.6_{-0.4}^{+0.3}$   & $4.2_{-0.4}^{+0.8 }$& $7.0_{-1.3}^{+0.9}$        & $29.1_{-0.5}^{+0.2}$     & 0.145$\pm$0.025        \\
0.9             & $0.7_{-0.4}^{+0.3}$   & $4.0_{-0.2}^{+0.9}$ & $6.5\pm1.3$        & $29.1_{-0.4}^{+0.3}$    & 0.14$\pm$0.025       \\ \hline

\end{tabular}
\end{table*}

\begin{figure*}
\includegraphics[width=177mm]{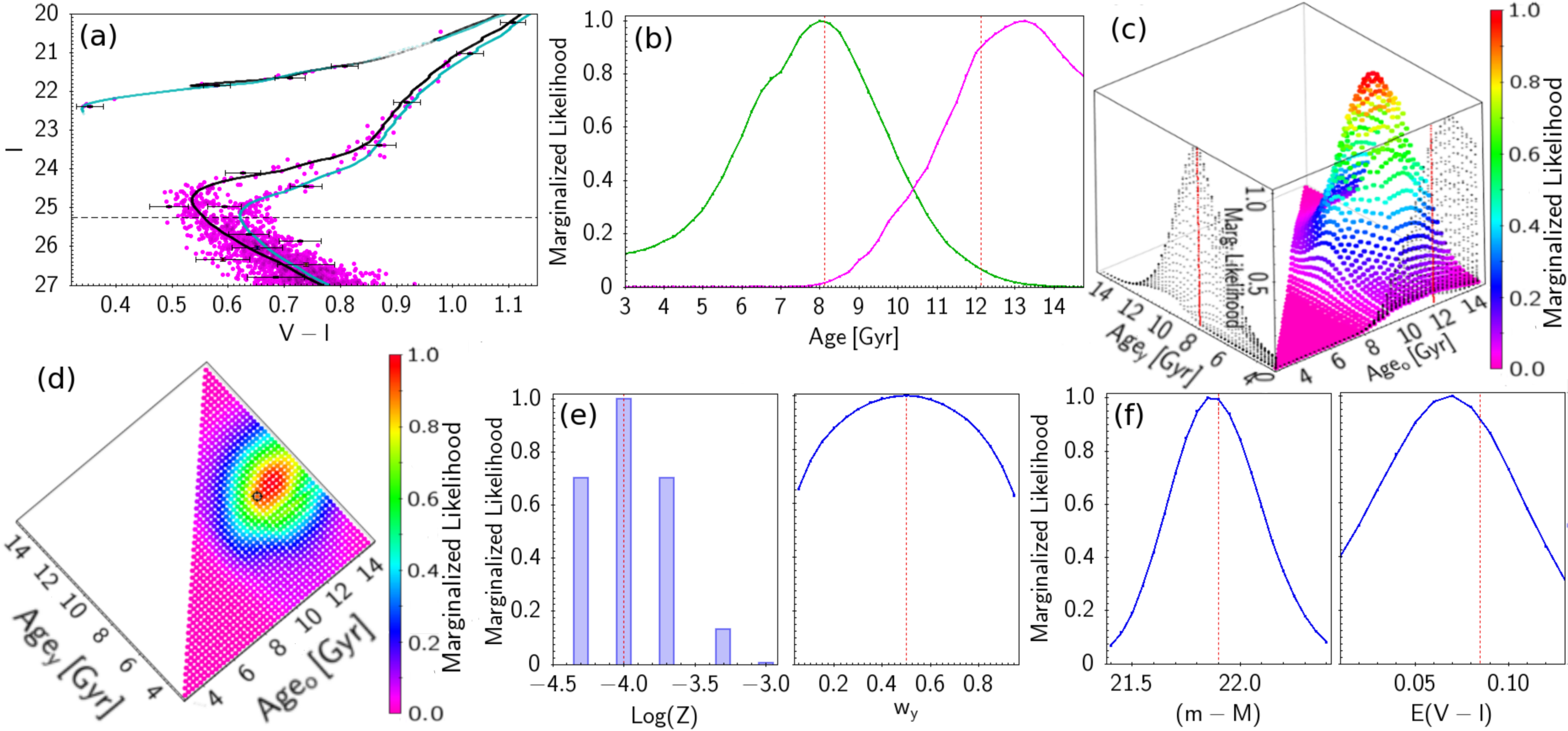}
\caption{(a) CMD of a mock galaxy with 2 old bursts (8.125 and 12.125 Gyr) for $w_{y,input} = 0.50$.
Input isochrones of 8.25 Gyr {\it (black)} and 12.0 Gyr {\it (cyan)} are shown.
(b) $\mathcal{L}_N^{marg}$ distribution for $t_y$ and $t_o$.
(c) 2D  $\mathcal{L}_N^{marg}$ distribution for $t_o$ and $t_y$. Each distribution is projected on a side and the input age is indicated by a {\it red line}.
(d) Same as (c) as a contour map, the input pair of ages is highlighted with a circle.  
(e,f) $\mathcal{L}_N^{marg}$ distributions for $Log\,Z$, $w_y$, $(m-M)$, and $E(V-I)$.
}
\label{fig4}
\end{figure*}

\begin{figure}
\includegraphics[width=84mm]{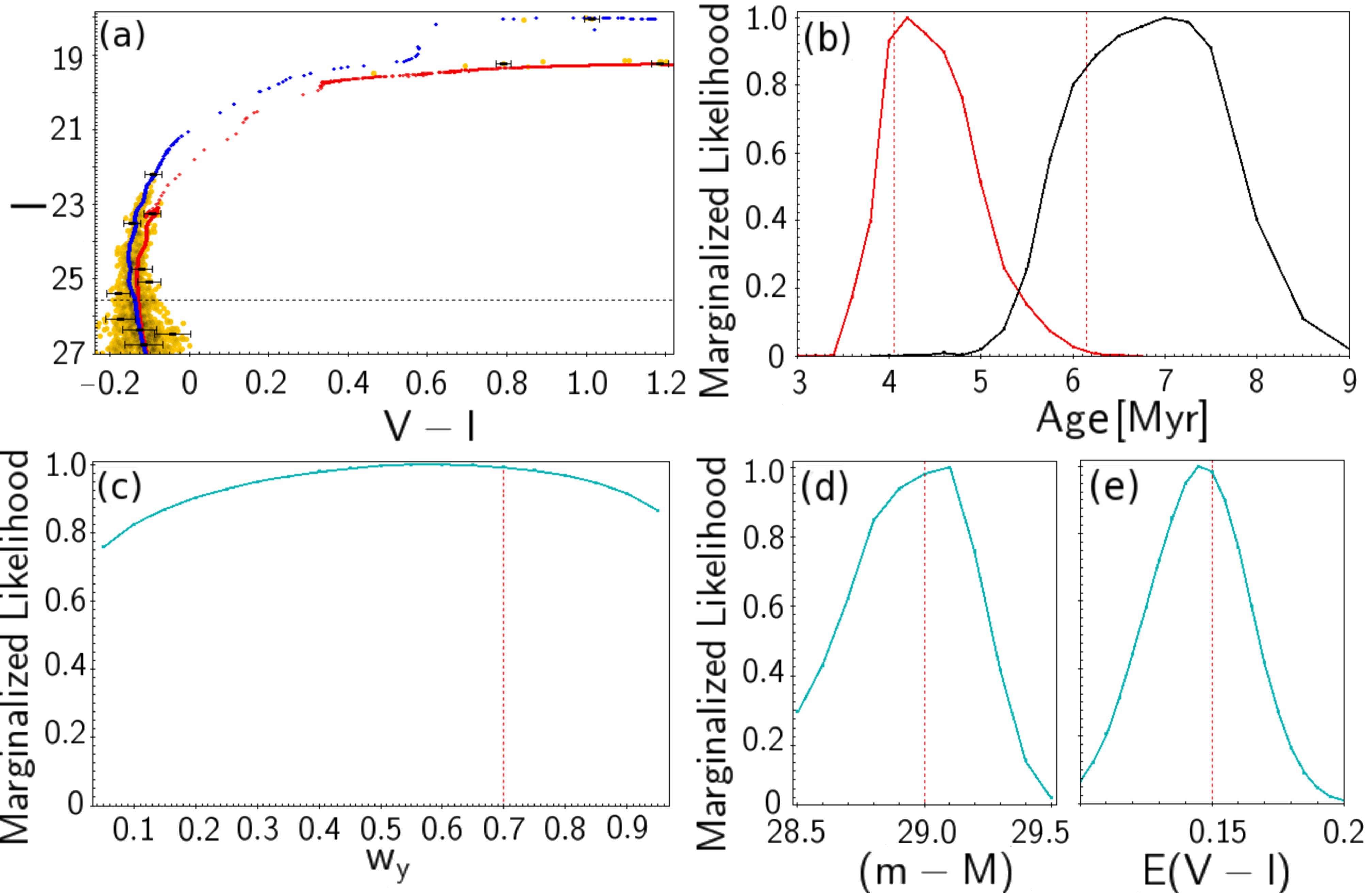}

\caption{
(a) CMD of the mock population with 2 young bursts (4.05 and 6.12 Myr) for $w_{y,input} = 0.70$.
The 4.2 and 7.0 Myr isochrones are shown in {\it blue} 
and {\it red}.
(b-e) $\mathcal{L}_N^{marg}$ distributions for $t_y, t_o,w_y,(m-M)$, and $E(V-I)$.}
\label{fig5rev}
\end{figure}

\begin{figure}
\includegraphics[width=84mm]{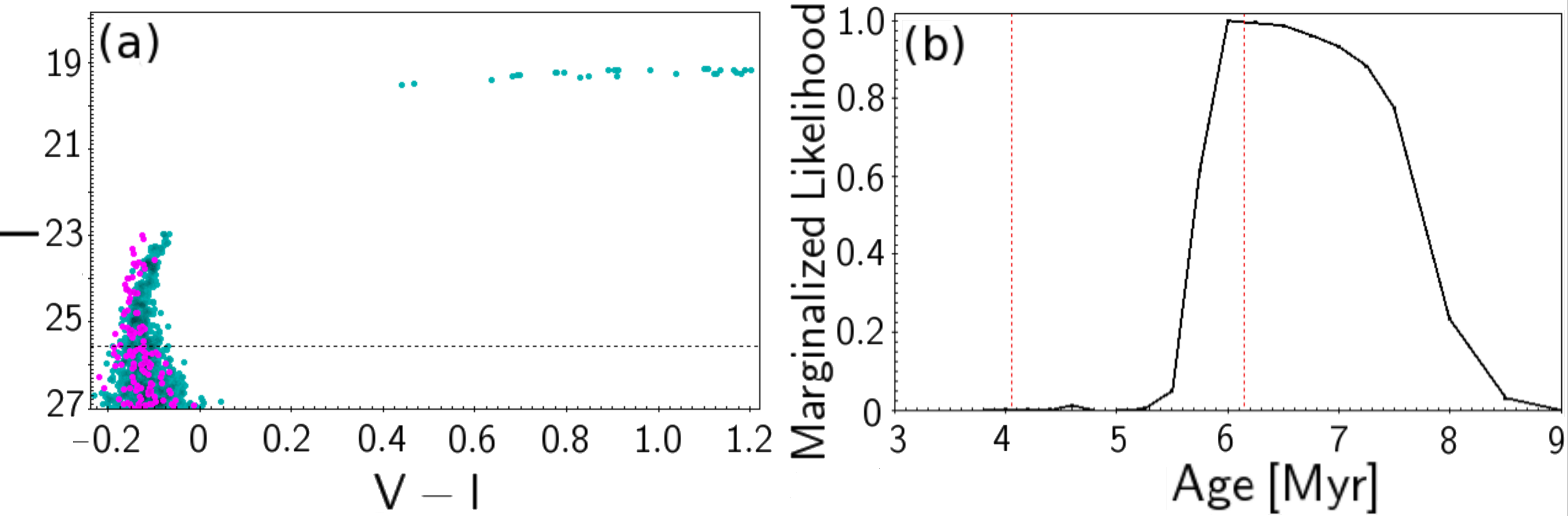}

\caption{
(a) CMD of the mock population with 2 young bursts (4.05 and 6.12 Myr) for $w_{y,input} = 0.10$. The oldest component is shown in {\it cyan} and the youngest one in {\it purple}. 
(b) $\mathcal{L}_N^{marg}$ distribution for the age. Only the oldest burst is recovered. The vertical lines indicate the age input values.
}
\label{fig6rev2}
\end{figure}

In this section we explore how our parameter estimator handles double stellar populations.
We consider a $Log \,Z= -4$ mock galaxy in which $w_o = 0.5$ of the stars were formed in an initial 
burst\footnote{
As in \S\ref{mock1}, there are no isochrones in our grid for age $t_o$ and $t_y$.
For the older population we add in equal amounts stars from the 12 and 12.25 Gyr isochrones, and for the younger population we add stars from the 8 and 8.25 Gyr isochrones.}
of age $t_o = 12.125$ Gyr and $w_y = 0.5$ of the stars in a younger burst happening at  $t_y = 8.125$ Gyr, seen at a $(m-M) = 21.9$ with $E(V-I) = 0.085$.
Photometric uncertainties were added according to Eq.~(\ref{eq13}), adding in quadrature a systematic uncertainty of 0.02.
The total sample amounts to 1600 stars, the CMD is shown in Fig.~\ref{fig4}a.
Only the 394 stars above $I = 25.5$ were taken in consideration for the analysis.
For each burst we estimate its age and weight, along with their common $(m-M)$ and $E(V-I)$, 
assuming as flat priors $21.4<(m-M)<22.4$, $0.03<E(V-I)<0.13$,  and $Log \,Z =\,$-3, -3.3, -3.7, -4, and -4.3. 
Figs.~\ref{fig4}b-d show the resulting $\mathcal{L}_N^{marg}$  distributions for $t_y$ and $t_o$ as 2D and 3D representations. 
The distribution for $t_o$ looks truncated at the high age end in Fig.~\ref{fig4}b because the 14.75 Gyr isochrone ($\sim 1$ Gyr older than the universe) is the oldest one in our grid.
Figs.~\ref{fig4}e-f show the $\mathcal{L}_N^{marg}$ distributions for $Log\,Z$, $w_y$, $(m-M)$, and $E(V-I)$.

Our parameter estimates closely correspond to the input values listed in Table~\ref{tab2}.
We remark that the confidence interval for $t_o$ is wider than for $t_y$, even though in relative terms they are similar
($\sim 23$\% for $\mathcal{L}_N^{marg}$).
Two factors contribute to broaden the confidence interval for $t_o$: {\it (1)} given the position of the younger starburst in the CMD, its MS stars contaminate 
the MSTO of the older starburst, while the opposite does not occur (this is the rule for double stellar populations, unless the photometric precision is outstanding), 
and {\it (2)} the MSTO stars of the younger burst are 0.5 mag brighter and have smaller photometric errors than the older burst stars.
Both $t_o$ and $t_y$ have larger confidence intervals as compared to the age of the simple population discussed in \S\ref{mock1}.
This is a consequence of the fact that stars from both bursts share common regions in the CMD, contaminating and broadening critical zones in this diagram with
respect to the simple population case.
Even though the estimated $w_y$ is close to the input value, the $\mathcal{L}_N^{marg}$ distribution for this
parameter in Fig.~\ref{fig4}f is wide, flat to a degree, and non-Gaussian, hence the confidence interval for $w_y$ is ill defined. 
The probability of a single stellar population never falls below 60\%, hence, these distributions can not be used to rule out this possibility.
If the double burst nature of a population is known a priori, 
our method recovers successfully the age of each burst.
The accuracy of the $Log\,Z$, $(m-M)$, and $E(V-I)$ estimates are similar to those in \S\ref{mock1}.

To further explore our parameter estimation power we generated several mock galaxies with identical parameters to the previous one but varying $w_y$ as indicated in Table~\ref{tab3}, where we also list our
results.
As expected, the confidence intervals in $t_o$ and $t_y$ tend to increase
as the weight of the concerned burst becomes smaller. 
This is manly due to the pollution by stars from the more numerous burst.  
In Table~\ref{tab3} there are two cases in which one of the bursts contributes 90\% of the stars.
We can consider these two cases as simple populations with a 10\% of contamination of stars from the
other burst, and compare with the results for the simple population in \S\ref{mock1}.
The estimated age for the burst with 90\% of the stars has a similar relative uncertainty
($\sim 21$\% for $\mathcal{L}_N^{marg}$)
to that of the simple population.
The age of the contaminating bursts (contributing with 10\% of the stars) is recovered
to some extent in both cases, with a relative uncertainty reaching values as high as $\sim 42\%$.
As expected, the accuracy of our age estimates increases with the relative strength of the burst.

\subsubsection{Two young stellar populations}\label{mock3}

Here we repeat the experiment performed in the last part of \S\ref{mock2} but for a $Log\,Z=-2.1\pm0.04$ mock 
galaxy containing two young stellar populations with ages $t_o=6.12$   and $t_y=4.05$ Myr 
$(m-M)=29.0$, and $E(V-I)=0.15$. The input weight $w_y$ assigned to the younger
population is listed in Table~\ref{tab4} together with our results.
The galaxy contains 1200 stars but only the 403 stars brighter than $I = 25.25$ were used.
As priors we use $28.5<(m-M)<29.5$, $0.10<E(V-I)<0.2$. We assume a prior knowledge of $Log\,Z=-2.1\pm{0.05}$, and marginalize over the isochrone 
sets with $-2.4<Log\,Z<-1.85$, covering the metallicity domain at the $\pm5\sigma$ level (cf.\S\ref{ParametrosYPriors}).
Figs.~\ref{fig5rev}a-e show the CMD, and the ($t_y, t_o, m-M, E(V-I), w_y$)
$\mathcal{L}_N^{marg}$ distributions for $w_{y,input} = 0.70$.
Figs.~\ref{fig6rev2}a-b show the CMD, and the $(t_y, t_o)$ $\mathcal{L}_N^{marg}$ distributions for $w_{y,input} = 0.10$.
In the latter case only the oldest burst is recovered.
The recovery of the parameters for the double young burst galaxy shows similar dependence on $w_y$ as the double old burst galaxy discussed in \S\ref{mock2b}.
Again, the $\mathcal{L}_N$ distributions for $w_y$ are not sharp enough as to discard a single burst nature
for this population.

\section{Applications}

\subsection{Star Clusters in the LMC and SMC}\label{LMC}

One of the advantages of our parameter estimator is that it can be used in a fast and objective manner to
characterize resolved stellar groups or clusters in nearby galaxies.
Using this procedure \cite{Bitsakis2017, Bitsakis2018} determined the age of 4850 star clusters 
in the LMC and 1319 star clusters in the SMC, respectively.
The clusters were previously defined by these authors as spatial overdensities in the Magellanic Cloud Photometry Survey,\footnote{
Carried out by \cite{Zaritsky2002, Zaritsky2004} with the 1m diameter Swope Telescope at Las Campanas Observatory. The data comprise two photometric catalogs of the stars in the LMC and the SMC in the Johnson $BVI$ and Gunn $i$ bands, corrected for internal and Galactic extinction.} using a new statistical analysis.
The resulting candidate clusters are cleaned from contaminating stars using the following probabilistic algorithm. 
They draw a box that encloses all the stars defining a candidate cluster, and a similar box in a neighbouring field,
and build the CMD of the stars inside each box.
Both CMDs are equally binned along their axes, and the number of stars inside each bin is counted.
Following \cite{Mighell1996}, the membership probability of the stars in a bin of the candidate cluster CMD is given by
\begin{equation}\label{eq14}
p_{memb}=1-\frac{N_{*,field}}{N_{*,cluster}}\cdot\frac{A_{cluster}}{A_{field}},
\end{equation}
where $N_{*,cluster}$ and $N_{*,field}$ are the number of stars in the cluster and field bins,
and ${A_{cluster}}$ and ${A_{field}}$ are the areas of the cluster and field boxes.

To quicken their analysis, \cite{Bitsakis2017, Bitsakis2018} dated all the star clusters assuming a single stellar population.
In this section we study in detail three LMC clusters from \cite{Bitsakis2017}: IR1-1959, NUV-1781,
and IR1-297.
For the first one we use a single stellar population in our analysis, for the other two, a double stellar
population.
Considering these examples is important since the CMD diagrams of these clusters exhibit more noise 
and scatter than the HST CMD of UFDGs analyzed in the next section, resulting in a more challenging parameter estimation.
Examples of two clusters in the SMC are shown by \cite{Bitsakis2018}.
For the LMC clusters we have prior knowledge of $Z, (m-M)$, and $E(B-V)$. We adopt $Log\,Z=-2.04\pm0.03$ and $(m-M) =18.48\pm0.05$, derived from Cepheids studies in the LMC by \cite{Keller2006}, and \citep{Walker2012}, respectively. By employing a set of isochrones with $Log\,Z=-2.22,-2.10,-2.00,-2.9,-2.85,-1.85$, we cover the metallicity domain at a $\pm 5\sigma$ level. The uncertainty on  $(m-M)$ was added in quadrature to the y-axis uncertainties in the CMD. Since the photometry of the stars has been corrected for extinction, we assume $E(B-V)=0$ and add in quadrature its 0.125 mag uncertainty to the x-axis uncertainties in the CMD \citep{Zaritsky2004}. Finally, an overall systematic uncertainty of 0.02 mag is added in quadrature to both axis. Then, our problem reduces to estimating the age $t$ of the stellar populations (cf.\S\ref{ParametrosYPriors}).

\begin{figure*}
\includegraphics[width=177mm]{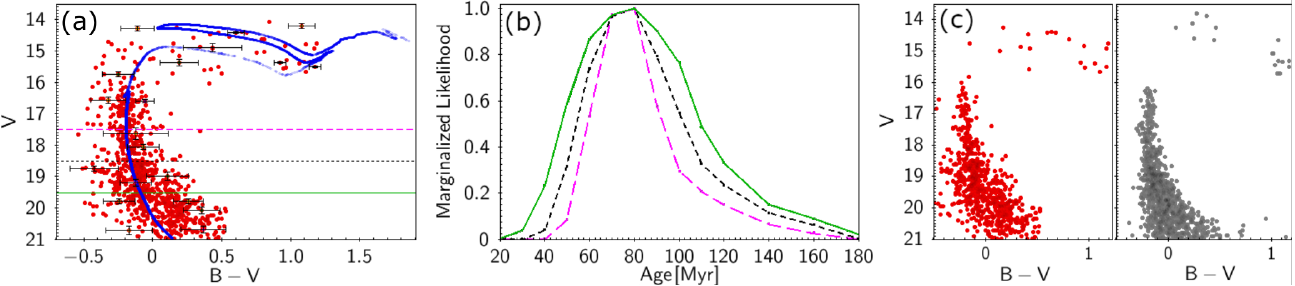}
\caption{(a) CMD of the cluster IR1-1959. Photometric errors are shown for some stars. The 80 Myr isochrone from the estimated age is shown in {\it blue}. 
(b) The age $\mathcal{L}_N^{marg}$ distributions for the three different subsamples defined by the $V$ magnitude cuts shown as horizontal lines in the top panel.
(c) {\it left:} Observed CMD.
(c) {\it right:} 80 Myr simple stellar population mock CMD generated as described in \S3.
}
\label{fig5} 
\end{figure*}

\begin{figure*}
\includegraphics[width=177mm]{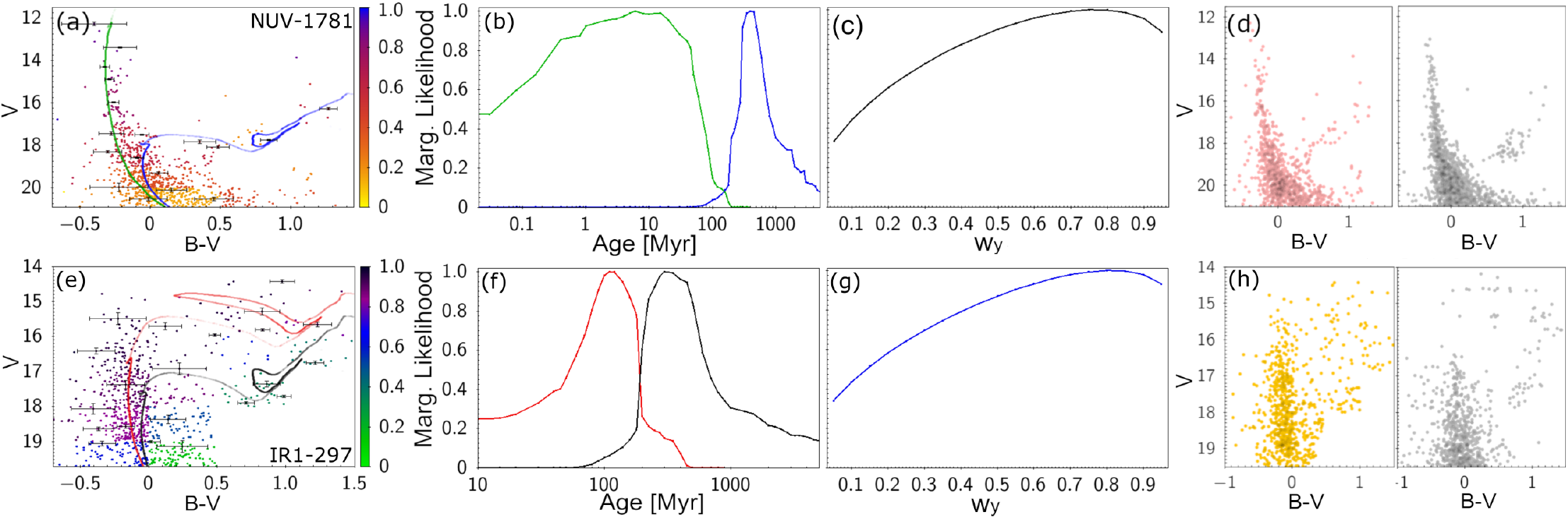}

\caption{
(a) CMD of LMC NUV-1781. Photometric errors are shown for some stars.
The isochrones from age estimates $t_y = 6$ Myr {\it (green)} and $t_o = 400$ Myr {\it (blue)} are shown.
(b) $\mathcal{L}_N^{marg}$ distribution functions for the age of the two stellar populations. 
(c) $\mathcal{L}_N^{marg}$ distribution for $w_y$. 
(d) Observed (LHS) and mock (RHS) CMD for the two populations.
\newline
(e) CMD of LMC IR1-297. Photometric errors are shown for some stars. The isochrones from age estimates  $t_y = 110$ Myr {\it (red)} and $t_o = 300$ Myr {\it (black)} are shown.
(f) $\mathcal{L}_N^{marg}$ distribution functions for the age of the two stellar populations.
(g) $\mathcal{L}_N^{marg}$ distribution for $w_y$. 
(h) Observed (LHS) and mock (RHS) CMD for the two populations.
}
\label{fig6}
\end{figure*}

\subsubsection{LMC IR1-1959 as a single stellar population}\label{LMCsingle}

Fig.~\ref{fig5}a shows the $(B-V,V)$ CMD for 700 stars in the cluster LMC IR1-1959, $\sim 30$ of them above the MSTO. 
The $\mathcal{L}^{marg}_N$ distributions for $t$ shown in 
Fig.~\ref{fig5}b 
were obtained using the $V$ magnitude cuts indicated in the
Fig.~\ref{fig5}a.
The three subsamples lead to the same age estimate, but the distributions broaden as we include fainter stars. For the 
subsample of brighter stars,
$t = 80^{+18}_{20}$ Myr. 
The RHS panel in Fig.~\ref{fig5}c
shows an 80 Myr simple stellar population mock CMD generated as described in \S3 next to the observed CMD (left hand side (LHS) panel).
The 60, 80, and 100 Myr isochrones are shown in 
Fig.~\ref{fig5}a.

\subsubsection{LMC NUV-1781 as a double stellar population} 

The CMD of LMC NUV-1781 shown in Fig.~\ref{fig6}a resembles that of a cluster with two stellar populations. This cluster has been cleaned of contaminating stars, however, there is still the possibility that remnant field stars may be defining a false secondary population, without being physically associated with the cluster \citep{Cabrera2016}. To account for this possibility, we adopt the approach of \cite{Bitsakis2017, Bitsakis2018}. The probability that each star is drawn from an isochrone (Eq. \ref{eq4}) is weighted by the membership probability: $p_{memb}\times p(s_k|h_i)$.

From the $\mathcal{L}_N^{marg}$ distribution in Fig.~\ref{fig6}b, LMC NUV-1781 contains a young population of age $t_y = 6_{-5.96}^{+58}$ Myr, and an older population of age $t_o = 400_{-150}^{+420}$ Myr.
The $t_y$ confidence interval ranges from 0.04 to 64 Myr. 
The similarity of the isochrones in this age range, the poorly defined upper MS (due to the scarcity of stars), and the photometric uncertainties, are responsible of the width of this confidence interval.
In contrast, the $\mathcal{L}_N^{marg}$ distribution for $t_o$ in Fig.~\ref{fig6}b results in a narrower, in relative terms, confidence interval for $t_o$, ranging from 250 to 820 Myr.
The presence of RGB stars in the older burst (in contrast with the lack of them in the younger one) provides a second feature that drives our matching isochrone algorithm to obtain a more precise estimate for $t_o$ than for $t_y$.
The $\mathcal{L}_N^{marg}$ distribution for $w_y$ 
(Fig.~\ref{fig6}c) indicates that the younger burst accounts for $75^{+25}_{-30}\%$ of the stars. This
result is consistent with the presence of a single stellar population in this cluster.
Higher precision photometry is needed for a more precise age estimate, as well as for supporting the double population hypothesis..
The RHS panel of Fig.~\ref{fig6}d shows a mock CMD for this population next to the observed one (LHS).

\subsubsection{LMC IR1-297 as a double stellar population}

The CMD of LMC IR1-297 is presented in Fig.~\ref{fig6}e. 
The $t_y$, $t_o$ and $w_y$ $\mathcal{L}_N^{marg}$ distributions (Fig.~\ref{fig6}f, g)  
were obtained in the same way as for LMC NUV-1781. From these distributions, we estimate $t_y = 110_{-60}^{+80}$ and $t_o = 300_{-110}^{+370}$ Myr. 
The $\mathcal{L}_N$ distributions for $w_y$ indicate that the younger burst may account for up to $\sim 80^{+20}_{-40}$\% of the cluster population. As for NUV-1781, higher precision photometry is required for a more precise age determination and to exclude or validate the double population hypothesis for this cluster.
The RHS panel of Fig.~\ref{fig6}h shows a mock CMD for this population next to the observed one (LHS).

\subsection{UFDGs}\label{UFDG}

\begin{figure*}
\centering
\includegraphics[width=177mm]{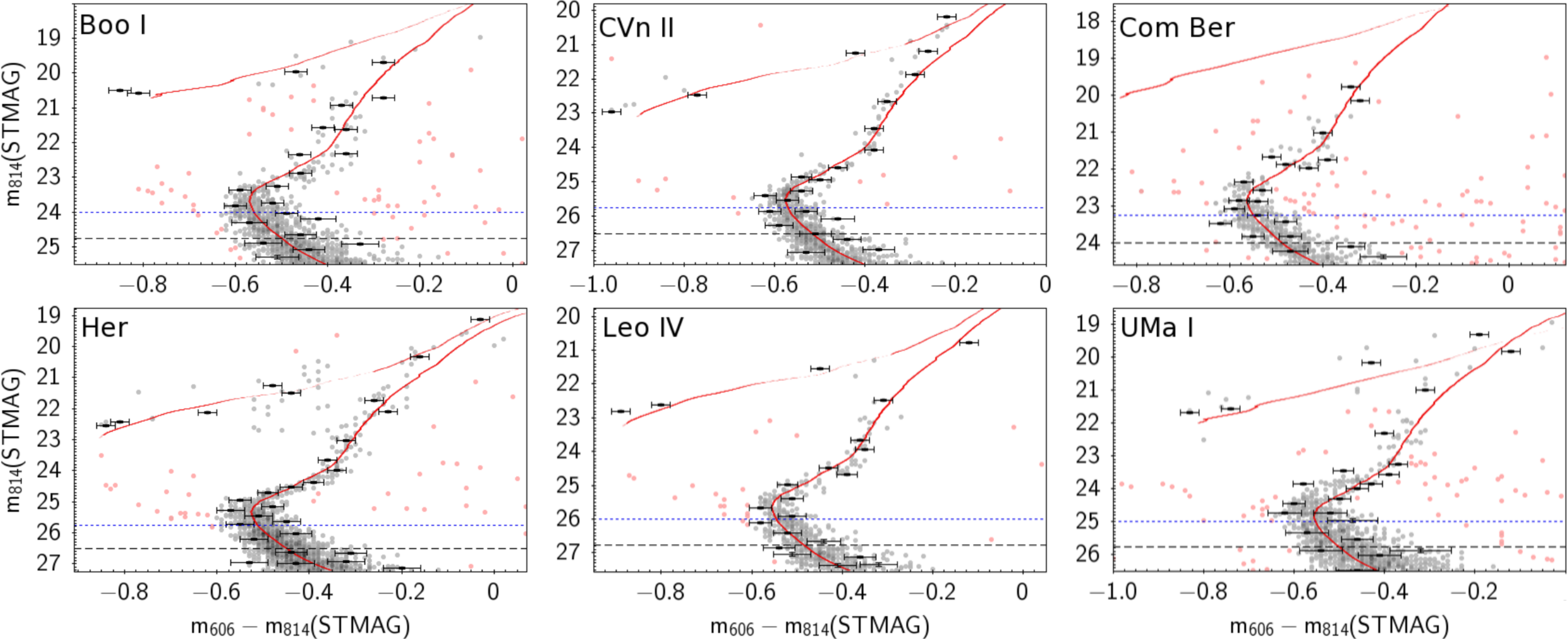}
\caption{
CMDs of the UFDGs under analysis. 
The horizontal lines mark two different subsampling trimming cuts in $m_{814}$.
The isochrones from the age estimates from the $\mathcal{L}_N^{marg}$ distributions are shown.
}
\label{fig7}
\end{figure*}

\begin{figure*}
\centering
\includegraphics[width=177mm]{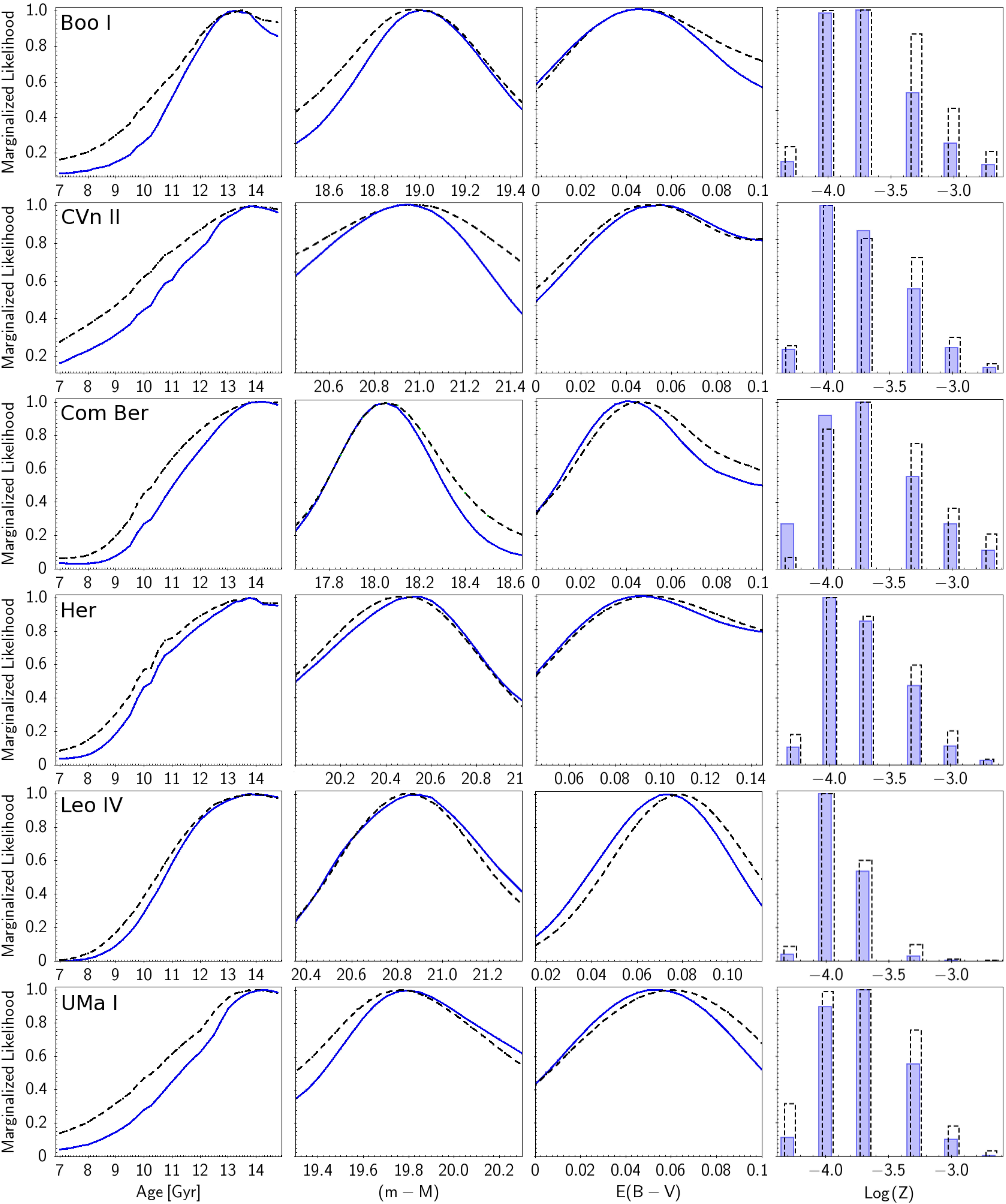}
\caption{
$\mathcal{L}_N^{marg}$ distributions for $t, (m-M)$, $E(B-V)$, and $log\,Z$. Two different subsamples were defined for each UFDG by trimming the sample below the colour coded $m_{814}$ magnitude cuts shown in Fig.~\ref{fig7}.
}

\label{fig8}
\end{figure*}

\begin{figure}
\centering
\includegraphics[width=67mm]{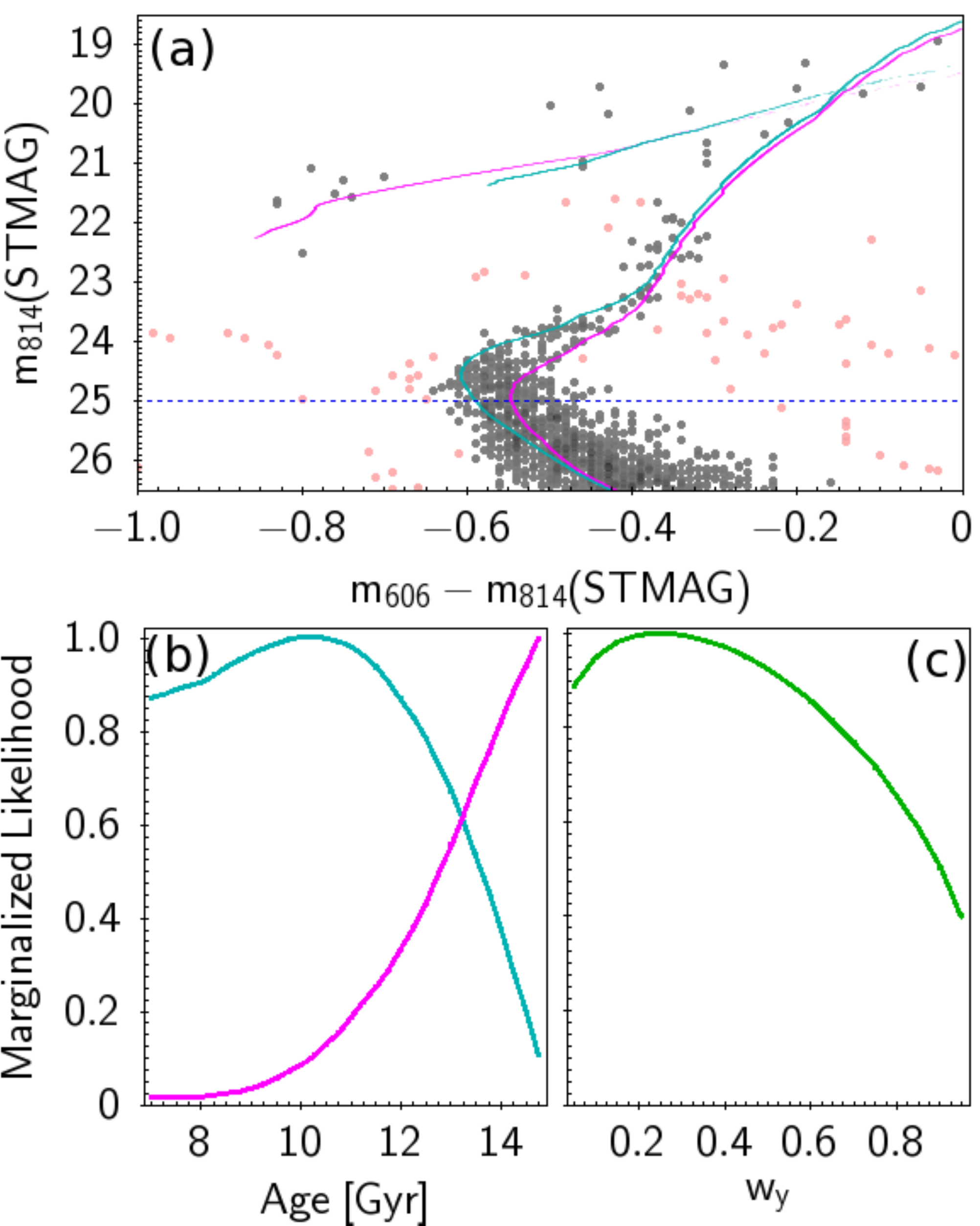}
\caption{
Two stellar populations in UMa I.
$(a)$ Isochrones from age estimates 11.5 Gyr ({\it blue}) and 14.75 Gyr ({\it magenta}) are shown.
We remark the width of the SGB. Only stars above the doted horizontal line were considered in the analysis. 
$(b)$ $\mathcal{L}_N^{marg}$ distributions for $t_y$ and $t_o$.
$(c)$ $\mathcal{L}_N^{marg}$ distribution for $w_y$.
}
\label{fig9}
\end{figure}

In this section we discuss our analysis of six UFDGs, satellites of the Milky Way: Bootes I (Boo I), Canes Venatici II (CVn II), Coma Berenices (Com Ber), Hercules (Her), Leo IV, and Ursa Major I (UMa I).
Deep optical images of these galaxies were obtained with the Advanced Camera for Surveys (ACS) on the HST using filters F606w and F814w (GO-12549; PI: T.M. Brown). 
The photometric data reduction was carried out by \cite{Brown2012, Brown2014}, who also obtained metallicity distributions from spectroscopic observations.
The photometry includes stars at least three magnitudes below the MSTO.
Contaminating stars and background galaxies were cleaned by rejecting sources with poor photometry, and with no typical stellar profiles on their point spread functions.
For a detailed description of the observations and data reduction we refer the reader to \cite{Brown2012, Brown2014}.

The ACS photometry of the identified UFDG stellar sources with photometric uncertainties is provided in the STMAG system as $m_{606}$ and $m_{814}$, where
\begin{equation}\label{stmag}
m_\lambda = -2.5 \times log\,f_\lambda - 21.1,
\end{equation}
with an overall systematic uncertainty of 0.02 mag. This amount is added in quadrature to the photometric uncertainties.
We computed a grid of isochrones in the STMAG system spanning from 7 to 14.75 Gyr, in 0.25 Gyr steps, for $log\,Z = -3.3, -3.7, -4$, and $-4.3$, which bracket the value of $log\,Z\sim -4$ for UFDGS from previous works \citep{Brown2014,Kirby2008}.

For each UFDG we then estimate the parameters $t, Z,(m-M)$, and $E(B-V)$.
The extinction correction in STMAG for cool stars was obtained from \citet{Bedin2005}.

Fig.~\ref{fig7} shows the CMDs for the six UFDGs.\footnote{
Stars marked in  pink in Fig.~\ref{fig7} were removed, in order to clean contaminating outlier sources (such as foreground stars or background galaxies) that are non-modellable points by the isochrone set in consideration. These stars carry a very small $p(s_k|h_{i})$, which may be critically risky, in some cases, since a single star with $p(s_k|h_{i})=0$  may collapse the whole likelihood during the multiplication process of Eq.~(\ref{eq8}), as noted by \cite{Tolstoy1996}.
}
Two different subsampling trimming cuts in $m_{814}$ are used (cf. \S\ref{trimming}).
Fig.~\ref{fig8} shows two $\mathcal{L}_N^{marg}$ likelihood distributions for each parameter for each galaxy, corresponding to the two cuts in $m_{814}$.
The estimated values of the parameters for each UFDG are listed in Table \ref{tab5}, where for comparison
we also list previous determinations by other authors. 
The estimates from the marginalized distributions $\mathcal{L}_N^{marg}$ result in an age between 13.25 and 14.25 Gyr, with a relative uncertainty of about 21\%. 
These estimates include only the lower limit for the confidence interval. This is because our isochrones do not extend past 14.75 Gyr, which is not enough to establish the upper limit
for the confidence interval.
Given the latest estimate of the age of the universe of 13.8 Gyr \citep{Planck2016}, this is not a 
concern for objects as old as this set of UFDGs. 
Fig.~\ref{fig7} shows the most likely isochrone in the CMD for each UFDG, including the low age confidence boundary isochrone. The high age confidence boundary isochrone is shown only for Boo I.
The accuracy of our estimates of $log\,Z, (m-M)$, and $E(m_{606}-m_{814})$ from $\mathcal{L}_N^{marg}$
are $\sim 0.5$, $\sim 0.4$ mag, and $\sim 0.04$ mag, respectively.
Taking into account the confidence intervals, these results are in line with the values reported by the previous studies listed in Table \ref{tab5}.
Our results support a scenario in which the star formation in these UFDGs occurred as early as redshift $z\sim6$.

Our age estimates are consistent with those of \cite{Brown2014}, especially for Boo I, CVn II and Com Ber, for  which our estimates are within the reported confidence intervals.
Our results are also in good agreement with \cite{Weisz2014} for the case of Her, and Leo IV. 
For the case of CVn II, there is a discrepancy between the estimated ages by \cite{Brown2014} and \cite{Weisz2014}, the latter being significantly younger.
Our result supports the age determined by \cite{Brown2014}. 
The younger age for CVn II by \cite{Weisz2014} could be due to their lower quality photometric data,
which is remarkably noisier about the MSTO, as already discussed by \cite{Brown2014}. 

The comparison between the age estimates by \cite{Brown2014} and our work is particularly relevant, considering that both works rely on the same CMDs.
In general, the age estimates are consistent, but ours present larger uncertainties.  
Part of the differences between their results and ours is due to the different methods, with a non-negligible contribution from the different set of isochrones.  
Whereas our isochrones are computed in steps of 0.25 Gyr, the step in their grid is 0.1 Gyr. 
Their isochrones follow the proportions of the metallicity distribution from their spectroscopic data, using a grid with a fine step of 0.2 dex in [Fe/H], while our isochrones are limited to discrete published values of $Z$. 
Moreover, unlike our isochrones, theirs include binary systems, and thus they may be more realistic since binary stars in UFDGS may reach up to 50\% \citep{Geha2013}.
Even though we do not reach the accuracy of the work by \cite{Brown2014}, the goal of our technique is focused on estimating {\it simultaneously} a set of multiple parameters using {\it only} stellar photometry. 
Age determinations by procedures like that of \cite{Brown2014} are preceded by several steps, e.g., measuring the metallicity, and estimating $(m-M)$ and $E(m_{606}-m_{814})$. 
The advantage of our approach is the fast and simultaneous multiple parameter estimation, intended for automated analysis of massive data sets.

\subsubsection{Searching for double bursts in UFDGs}

We searched for a possible second stellar population in the UFDGs of our sample.
From our $\mathcal{L}_N$ distributions there is a hint of a possible double population only for UMa I (Fig.~\ref{fig9}b,c),
but they do not rule out the hypothesis of a single burst. 
Considering UMa I as a double stellar population, the most likely isochrones describe two possible SGBs in the CMD of this galaxy, 
as seen in Fig.~\ref{fig9}a. The $\mathcal{L}_N^{marg}$ distributions lead to ages of $t_y=10.0^{+3.5}_{-5.0}$ Gyr for the younger population and $t_o=14.75_{-2.0}$ Gyr for the older one, 
with a relative weight for the younger population of $w_y=0.25^{+0.50}_{-0.25}$.
Taking into account the uncertainties, there is not conflict between $t_o$ and the age of the universe. 
These estimates are in good agreement with those by \cite{Brown2014}, $t_y=11.6$ Gyr, $t_o=14.1$ Gyr, and $w_y=0.55$. 
Considering the overlap of the $\mathcal{L}$ distributions for $t_y$ and $t_o$ in Fig.~\ref{fig9}a, we cannot rule out
a scenario of continuous star formation starting $\sim 13.5$ Gyr ago and lasting for $\sim 1.5$ Gyr until $\sim 12$ Gyr ago.
Nevertheless, support for the double population nature of UMa I was found from its metal content by \cite{Webster2015}. 
\cite{Webster2015} compared the observed and the theoretical metallicity distributions of the UFDGs in our sample.
For UMa I, they explored two different enrichment scenarios. 
In the first scenario, an instantaneous burst occurs 14.1 Gyr ago with a extremely poor metallicity.
As the stars evolve, they enrich the medium from which a second population forms 2.5 Gyr later \citep[using the parameters determined by][]{Brown2014}.
In the second scenario, stars form continuously for $\sim 0.1$ Gyr, enriching the medium as they age.
They conclude that the observed distribution of metals in UMa I favours the first scenario  with a double burst.

For the other UFDGs in our sample, Boo I, CVn II, Com Ber, Her, and Leo IV we do not find significant evidence of the presence of a 
second population.
Either the estimates of $t_y$ and $t_o$ are so close within each other inside the confidence interval, or
the fraction of stars attributed to one of the bursts is too small ($\leq 15$\%).
\cite{Webster2015} find that the metallicity distributions of these five galaxies are reproduced better by a single continuous star 
formation model than by a double population.
For Leo IV, their results are not very conclusive due to the scarcity of stars.

\begin{table*}
\caption{Physical parameters of the UFDGs.}
\label{tab5}
\begin{tabular}{|l|l|l|l|l|l|}
\hline 
UFDG           & Work                   & $t$ (Gyr)                 & $log(Z)$                              & $(m-M)$                      & $E(B-V)$                                      \\ \hline
Bootes I       & Present     & 13.25$_{-2.25}$    & $-3.7_{-0.5}^{+0.4}$   & 19.0$\pm0.35$               & 0.05$\pm0.05$    \\

               & \cite{Brown2014}       & 13.3$\pm$0.3$\pm\sim1$    & $-4.1\pm0.5$              & 19.11$\pm0.07\pm0.01$        & 0.04$\pm0.01$    \\ \smallskip
               & \cite{McConnachie2012} & -                         & $-4.32\pm0.11$             & 19.11$\pm0.08$               & 0.017            \\

Canes Venatici II & Present            & 13.75$_{-3.25}$           &  $-4_{-0.3}^{+0.7}$      & 20.9$\pm0.5$                & 0.05$_{-0.04}^{+0.06}$    \\
               & \cite{Brown2014}       & 13.6$\pm0.3\pm\sim1$      & $-4.7\pm0.6$              & 21.04$\pm0.06\pm0.01$        & 0.04$\pm0.01$    \\
               & \cite{Weisz2014}       & 10.0$^{+1.5}_{-1.1}$      & -                                & -                            & -                \\ \smallskip
               & \cite{McConnachie2012} & -                         & $-4.0\pm0.05$              & 21.02$\pm0.06$               & 0.010            \\
Coma Berenices & Present    & 14.0$_{-3.25}$           & $-3.7_{-0.6}^{+0.4}$      & 18.05$\pm0.35$                & 0.04$_{-0.03}^{+0.05}$    \\

                & \cite{Brown2014}       & 13.9$\pm0.3\pm\sim1$      & $-4.3\pm0.5$              & 18.08$\pm0.10\pm0.01$        & 0.04$\pm0.01$    \\ \smallskip
               & \cite{McConnachie2012} & -                         & $-4.4\pm0.05$             & 18.2$\pm0.20$                & 0.017            \\ 
Hercules       & Present     & 13.75$_{-3.25}$           & $-4_{-0.3}^{+0.7}$      & 20.5$\pm0.5$                & 0.09$_{-0.5}^{+0.7}$    \\
               & \cite{Brown2014}       & 13.1$\pm0.3\pm\sim1$      & $-4.4\pm0.4$              & 20.92$\pm0.05\pm0.01$        & 0.09$\pm0.01$    \\
               & \cite{Weisz2014}       & 13.0$^{+0.3}_{-2.8}$      & -                                & -                            & -                \\
               & \cite{McConnachie2012} & -                         & $-4.2\pm0.04$              & 20.6$\pm0.2$                 & 0.063            \\ \smallskip
               & \cite{Musella2012}     & -                         & $-4.1\pm0.15$              & 20.6$\pm0.2$                 & 0.09$\pm0.02$    \\
Leo IV        & Present    & 13.75$_{-3.0}$           & $-4_{-0.2}^{+0.4}$      & 20.9$\pm0.35$                & 0.07$\pm0.02$    \\
               & \cite{Brown2014}       & 13.1$\pm0.4\pm\sim1$      & $-4.3\pm0.6$              & 21.12$\pm0.07\pm0.01$        & 0.08$\pm0.01$    \\
               & \cite{Weisz2014}       & 11.7$^{+1.4}_{-4.2}$      & -                                & -                            & -                \\
               & \cite{McConnachie2012} & -                         & $-4.3\pm0.07$              & 20.94$\pm0.09$               & 0.026            \\ \smallskip
               & \cite{Moretti2009}     & -                         & $-4.1\pm0.1$               & 20.94$\pm0.07$               & 0.04$\pm0.01$    \\
Ursa Major I   & Present     & 14.0$_{-2.5}$           & $-3.7_{-0.5}^{+0.5}$      & 19.8$\pm0.5$                & 0.05$\pm0.05$    \\
               & \cite{Brown2014}       & 12.7$\pm0.3\pm\sim1$      & $-4.4\pm0.5$              & 20.10$\pm0.05\pm0.01$        & 0.05$\pm0.01$    \\
               & \cite{McConnachie2012} &  -                        & $-4.0\pm0.04$              & 19.93$\pm0.10$               & 0.020            \\
\hline                                                                                                
\end{tabular}
\begin{flushleft}
For our work we quote the $\mathcal{L}_N^{marg}$ estimates derived from the subsamples that include only the stars brighter than $\sim 0.5$ mag below the MSTO. 
$E(m_{606}-m_{814})$ was transformed to $E(B-V)$ following \cite{Bedin2005}.
The \cite{Brown2014} ages correspond to the mean age of two fits to the CMD.
The first uncertainty in $t$ and $(m-M)$ corresponds to the statistical uncertainty, the second value to a systematic uncertainty.
Their estimates were calibrated with the globular cluster M92, adopting an age of 13.2 Gyr, systematic uncertainty of $\sim 1$ Gyr, $(m-M)=14.62\pm0.01$, and $E(B-V)=0.023$.
Their metallicity is presented using the media and stdev from their reported metallicity distributions.
The \cite{Weisz2014} age indicate the epoch at which 70\% of the stellar mass was formed. Their uncertainties include the random and systematic uncertainties.
The \cite{McConnachie2012} values come from estimates by other authors using diverse techniques.
\end{flushleft} 
\end{table*}

\section{Conclusions}			

We present a fast code based on Bayesian inference with the main purpose of estimating the age, metallicity, distance modulus, and color excess of stellar populations from their CMD. 
We introduce a new method to estimate the ages of double (or multiple) stellar populations, as well as the stellar contribution of each star burst. 
The code evaluates the likelihood, $\mathcal{L}$, of each combination of parameters, producing a $\mathcal{L}$ distribution. 
We estimate the parameter values and their confidence intervals from the marginalized likelihood distributions, 
$\mathcal{L}^{marg}_N$.
The possibility of evaluating the parameter in this manner represents an alternative to the Monte Carlo approach, which requires a large number of calculations to compare several stochastically populated isochrones with the observed CMD \citep{Dolphin2002}.

We test our procedure on mock CMDs of simple stellar populations, recovering the full set of input parameters ($age, Z, m-M, E(V-I)$), and examined its performance on the age estimates depending on the number of stars on the CMD. 
We also explored the effects of the sample selection on the resulting parameter estimates.
We trimmed the observed and mock CMD samples by discarding the stars fainter than various cuts in
apparent magnitude, and found that keeping only the stars brighter than $\sim 0.5$ mag below the MSTO results in significantly lower uncertainties.
Including fainter stars is not inoffensive, enhancing the uncertainty on each of the estimated parameters.

We also study mock CMDs of double stellar populations.
We closely recover the correct values of each input parameter for each population, except in the cases of poorly populated bursts and of bursts of very similar age.
For the latter we recover the age of the most prominent burst only. 
The accuracy of the estimated age for each burst depends remarkably on its relative contribution to the number of stars in the double population. The age of the dominating burst is determined with higher accuracy.
Although the age estimate of each component of the double populations considered in this work
are satisfactory, the resulting $\mathcal{L}$ distributions of the weight of each population tend to be too wide, 
and non-Gaussian. They entail a low accuracy which do not allow us to discard the possibility of simple stellar populations. To establish the meanigfulness of the pair of estimated ages additional information favouring a double population is required.

We used our tool to analyze in detail three LMC star clusters from the \cite{Bitsakis2017} sample.
The CMD of cluster IR1-1959 clearly corresponds visually to a simple stellar population. 
The CMDs of NUV-1781 and IR1-297 show visual signs of double populations, an hypothesis that was tested in our analysis.
Using known priors for $Z$, $(m-M)$, and $E(B-V)$, for the LMC, we estimated the age of IR1-1959, and the age and relative fraction of each stellar population in NUV-1781 and IR1-297. Further work is required in order to establish the reliability of the number of star bursts that have occurred in these clusters.

Finally, we determined $t, Z, (m-M)$, and $E(B-V)$ for 6 UFDGs from their
HST ACS $(m_{606}-m_{814},m_{814})$ CMDs.
We obtained results consistent with previous works for all the parameters (Table \ref{tab5}).
Our results support the notion that UFDGs contain very ancient stars of first or second generation formed
in the early universe. The age of these galaxies is older than $\sim 13.5$ $Gyr$ and they are very metal poor, with $log\,Z \sim -4$. 
We searched for a possible second stellar population in the UFDGs in our sample, and found low signs 
of a double population for the case of UMa I, in agreement with previous work by other authors, although, our analysis also supports the single stellar burst hypothesis.

\section{Acknowledgments}

The authors thank Dr. T.M. Brown and his team for sharing the UFDG data used in this work. 
VRS thanks X. Hern\'andez and M. Cervi\~no for illuminating discussions during the early phases of this project. 
The authors acknowledge the thorough reading of the original manuscript by the reviewer, Nicolas Martin, as helpful in reaching a clearer and more complete final version.
The research in this paper is part of the PhD thesis of VHS in the Universidad Nacional Aut\'onoma de M\'exico graduate program in astrophysics.
He thanks the support from the Instituto de Radioastronom\'\i a and Astrof\'\i sica, its staff, and
CONACyT for the scholarship granted.
BCS acknowledges financial support through PAPIIT project IA103517 from DGAPA-UNAM.
GB acknowledges financial support through PAPIIT projects IG100115 and IG100319  from DGAPA-UNAM.
TB acknowledges support from the CONACyT Research Fellowships program. 
We gratefully acknowledge support from the program for basic research of CONACyT through grant number 252364.
The Figures in this article were produced using the TOPCAT graphics viewer and tabular data editor.

\bibliographystyle{mnras}


\bsp	
\label{lastpage}
\end{document}